\documentclass[preprint,numberedappendix,appendixfloats,deluxetable]{emulateapj}
\slugcomment{ApJ, Accepted}
\pdfoutput=1
\usepackage{graphicx,amsmath,natbib, hyperref,amsfonts}
\DeclareMathOperator*{\argmax}{arg\,max}
\DeclareMathOperator*{\argmin}{arg\,min}

\shorttitle{Reconnaissance of the HR 8799 Exosolar System II}
\begin{document}

\title{Reconnaissance of the HR 8799 Exosolar System II: Astrometry and Orbital Motion}

\author{L. Pueyo\altaffilmark{1}}
\author{R. Soummer\altaffilmark{1}}
\author{J. Hoffmann\altaffilmark{2}}
\author{R. Oppenheimer\altaffilmark{3}}
\author{J. R. Graham\altaffilmark{15}}
\author{N. Zimmerman\altaffilmark{13}}
\author{C. Zhai\altaffilmark{6}}
\author{J. K. Wallace\altaffilmark{6}}
\author{F. Vescelus\altaffilmark{6}}
\author{A. Veicht\altaffilmark{3}}
\author{G. Vasisht\altaffilmark{6}}
\author{T. Truong\altaffilmark{6}}
\author{A. Sivaramakrishnan\altaffilmark{ 1}}
\author{M. Shao\altaffilmark{6}}
\author{L. C. Roberts, Jr.\altaffilmark{6}}
\author{J. E. Roberts\altaffilmark{6}}
\author{E. Rice\altaffilmark{11, 1}}
\author{I. R. Parry\altaffilmark{10, 1}}
\author{R. Nilsson\altaffilmark{3}}
\author{S.  Luszcz-Cook \altaffilmark{3}}
\author{T. Lockhart\altaffilmark{6}}
\author{E. R. Ligon\altaffilmark{6}}
\author{D. King\altaffilmark{10}}
\author{S. Hinkley\altaffilmark{5} \altaffilmark{8}}
\author{L. Hillenbrand\altaffilmark{7}}
\author{D. Hale\altaffilmark{4}}
\author{R. Dekany\altaffilmark{4}}
\author{J. R. Crepp\altaffilmark{11}}
\author{E. Cady\altaffilmark{6}}
\author{R. Burruss\altaffilmark{5}}
\author{D. Brenner\altaffilmark{3}}
\author{C. Beichman\altaffilmark{4, 5}}
\author{C. Baranec\altaffilmark{14}}

\altaffiltext{1}{Space Telescope Science Institute, 3700 San Martin Drive, Baltimore, MD 21218 USA,  e-mail inquiries should be directed to pueyo@stsci.edu. A description of the contributions of each author can be found at \url{http://www.amnh.org/project1640}}
\altaffiltext{2}{Department of Physics and Astronomy, Johns Hopkins University, Baltimore, MD, USA}
\altaffiltext{3}{Astrophysics Department, American Museum of Natural History, Central Park West at 79th Street, New York, NY 10024 USA}
\altaffiltext{4}{Caltech Optical Observatories, California Institute of Technology, Pasadena, CA 91125 USA}
\altaffiltext{5}{NASA Exoplanet Science Institute, California Institute of Technology, Pasadena, CA 91125 USA}
\altaffiltext{6}{Jet Propulsion Laboratory, California Institute of Technology, 4800 Oak Grove Dr., Pasadena CA 91109 USA}
\altaffiltext{7}{Department of Astronomy, California Institute of Technology, 1200 E. California Blvd, MC 249-17, Pasadena, CA 91125 USA}
\altaffiltext{8}{Department of Physics and Astronomy, University of Exeter, Physics Building, Stocker Road, Exeter EX4 4QL}
\altaffiltext{10}{Institute of Astronomy, Cambridge University, Madingley Road, Cambridge CB3 0HA United Kingdom}
\altaffiltext{11}{Department of Engineering Science \& Physics, College of Staten Island, 2800 Victory Blvd., Staten Island, NY 10314 USA}
\altaffiltext{12}{Notre Dame University, Indiana USA}
\altaffiltext{13}{Max Planck Institute for Astronomy, Heidelberg, Germany}
\altaffiltext{14}{Institute for Astronomy, University of Hawai'i at Manoa, Hilo, HI 96720-2700, USA}
\altaffiltext{15}{Berkeley Astronomy Department, 601 Campbell Hall, University of California, Berkeley CA 94720-3411 USA}

%
%
%%>>>> Include a list of keywords after the abstract 
%\keywords{exo-planets, instrumentation, speckle calibration, high contrast imaging}

%begin{abstract}
%%
%FU Ori is the prototype of its class. Young systems outbursts that show accretion. Companion was discovered. Goal was to constrain the close environment of the companion using its spectrum. We used IFU to look at it: resolve the two objects. We used the first object as a calibrator. We measured the J and H band spectrum of the second. We see further evidence for extincation. We conclude something, or not
%end{abstract}

%%%%%%%%%%%%%%%%%%%%%%%%%%%%%%%%%%%%%%%%%%%%%%%%%%%%%%%%%%%%%
\begin{abstract}
We present an analysis of the orbital motion of the four sub-stellar objects orbiting HR8799. Our study relies on the published astrometric history of this system augmented with an epoch obtained with the Project 1640 coronagraph + Integral Field Spectrograph (IFS) installed at the Palomar Hale telescope. We first focus on the intricacies associated with astrometric estimation using the combination of an Extreme Adaptive Optics system (PALM-3000), a coronagraph and an IFS. We introduce two new algorithms. The first one retrieves the stellar focal plane position when the star is occulted by a coronagraphic stop. The second one yields precise astrometric and spectro-photometric estimates of faint point sources even when they are initially buried in the speckle noise. The second part of our paper is devoted to studying orbital motion in this system. In order to complement the orbital architectures discussed in the literature, we determine an ensemble of likely Keplerian orbits for HR8799bcde, using a Bayesian analysis with maximally vague priors regarding the overall configuration of the system. While the astrometric history is currently too scarce to formally rule out coplanarity, HR8799d appears to be misaligned with respect to the most likely planes of HR8799bce orbits. This misalignment is sufficient to question the strictly coplanar assumption made by various authors when identifying a Laplace resonance as a potential architecture. Finally, we establish a high likelihood that HR8799de have dynamical masses below $13 \; M_{Jup}$ using a loose dynamical survival argument based on geometric close encounters. We illustrate how future dynamical analyses will further constrain dynamical masses in the entire system. 
\end{abstract}

\maketitle 

\section{Introduction}\label{intro}

\subsection{Orbital motion: a key element of direct imaging surveys}

High-contrast imaging of nearby stars is a powerful tool to acquire novel insights regarding the architecture and formation history of planetary systems. Such observations are indeed sensitive to sub-stellar companions and faint planets in a separation regime ($\sim >10$AU) difficult to reach using indirect methods \citep{oh09,vcf09,cj11}. They also enable to survey the vicinity of young and adolescent stars \citep{2011PASP..123..692M,bkt10} and thus provide direct constraints on the early stages of planetary formation and evolution \citep{sb11,bcb10}. Over the past few years a handful of such objects have been directly imaged \citep{cld05,mmb08,kgc08,ljv08,lbc10,mzk10,ikm11}. Because their near-infrared radiation is readily available for characterization, their discovery has spurred numerous follow-up photometric \citep{qmk10,cbi11,gmm11,jcl12,2013A&A...549A..52E,2012ApJ...755L..34C,she12} and spectroscopic observations \citep{jbg10,bld10,bmk11,bmk11b,2013ApJ...768...24O}. This wealth of information, only available in favorable configurations in the case of exo-planets detected with indirect methods, has in turn inspired discussions regarding their underlying bulk physical properties and atmospheric chemistry \citep{2011ApJ...737...34M,bmk11,bmk11b,2012ApJ...754..135M}. Their unique loci in the separation vs. age plane combined with their rich observable astrophysical content makes directly imaged exo-planets very compelling comparative exo-planetology objects. Large observational programs, relying on new generation instruments, aimed at identifying more of such faint companions are currently underway or about to be started \citep{bfd08,mgp08,mks08,hoz11}.The observation underlying this paper were obtained using one of such instruments: the Project 1640 Integral Field Spectrograph \citep{hoz11} installed at the Palomar Hale Telescope behind the PALM 3000 Adaptive Optics system \citep{0004-637X-776-2-130}. 

Using an IFS one can reveal near-infrared spectroscopic features of sub-stellar companions to nearby stars, study their atmosphere and infer their bulk physical properties \citep{bmk11b,bmk11b,2013ApJ...768...24O,2013Sci...339.1398K,2013ApJ...779..153H}. However spectroscopic observations cannot fully address uncertainties in the mass-luminosity relationship of sub-stellar objects at young ages ($<100$ Myrs), since imaging does directly yield observables commensurate with dynamical masses. They are inferred by folding together estimated age (based on stellar indicators) and mass-luminosity relationships (based on evolutionary models), onto their observed photometric points. Calibrating this relationship at young ages is thus of the utmost importance. A key component of the upcoming large surveys will be to obtain, at least for a subset of the discovered objects, model independent dynamical mass estimates. For later type host stars this can be accomplished by obtaining three dimensional orbits that combine direct imaging and radial velocity observations \citep{cjf12}. For low mass ratio binaries with physical small separations this is achieved via direct astrometric monitoring over a full orbital period of the binary pair \cite{2009ApJ...706..328D,2010ApJ...711.1087K}. However for young sources with orbital periods $>10$ yrs and high mass ratio, it is very difficult to observe the gravitational influence of the companion on its host star. One of the most promising avenue to obtain dynamical masses for such young benchmark objects is to use the companion orbital motion to constrain the second order dynamical interaction between the various components in a multiple system \citep{fm10}, or a single planet and a circumstellar disk \citep{2012A&A...542A..41C,2013ApJ...775...56K}. In all cases, precise orbital characterization, and thus precise astrometry, is at the crux of this mass determination. Because all future direct imaging campaigns will rely on an Integral Field Spectrograph as the main survey camera, robust astrometric estimators with such instruments is of critical interest. The first goal of this paper is to introduce such a tool to the high-contrast imaging community. In Section \ref{sec:Data} we discuss how to retrieve not only the spectra but also the relative position of faint planets with respect to their host star, in the regime where they are buried under quasi-static speckles.\\ 

\subsection{Orbital motion in the HR8799 multiple system}

HR8799 is a nearby ($d\sim 30$ pc) young star ($30$ Myrs), which harbors a multiple planetary system, with four planets orbiting at separations ranging from $\sim 20 $ to $\sim 75$ AU \citep{mmb08,mzk10}. In a parent paper by \citet{2013ApJ...768...24O}, we reported near-infrared ($1-1.8 \; \mu m$) spectroscopic observations ($R \sim 40$) of the four planets in this system. Our results highlighted how the Spectral Energy Distributions of these objects are different from known brown dwarfs, and established their spectral diversity, in spite of having formed in the same circumstellar environment. These spectra are also sensitive to a variety of molecular opacities in the atmosphere of each planet, and will thus be the observational foundation of future theoretical work aimed at understanding their complex atmospheric chemistry. The HR8799 system is extremely interesting from a dynamical mass determination standpoint because of its high multiplicity. Since its discovery numerous epochs of this system have been reported \citep{hrk10, 2013A&A...549A..52E,cbi11,hci11,2012ApJ...755L..34C,gmm11,jbg10,lmd09,mmb08,mzk10,smb10,she12,2012ApJ...755L..34C}. This provides a finely sampled orbital coverage starting in 2008. Moreover \citet{Sou11} recently unravelled the three outermost planets in 1998 HST-NICMOS archival data, yielding a sufficiently large temporal baseline to constrain the eccentricity of the second innermost planet. Before the detection of HR8799e various authors considered the dynamical architecture of this system and suggested that the masses of HR8799bcd might be lower than estimated in the discovery paper \citep{mmb08} in order for the system to have remained stable over its lifetime \citep{mzk10,fm10}. \cite{2013A&A...549A..52E} recently combined the 1998 HST-NICMOS points with an early estimate for the orbit of HR8799e ($\sim 3$ yrs of temporal baseline) and suggested that indeed the dynamical masses of these planets lie around $\sim 7 \; M_{Jup}$. 

This paper reports the orbital position of HR8799bcde at our P1640 epoch, and then establishes the subspace of orbits allowable given the collection of epoch obtained over the past few years. We do so by resorting to a Bayesian analysis using Markov Chain Monte Carlo (MCMC). We carry out this work in a effort to complement recent orbit fitting efforts and dynamical investigations which assumed combinations of coplanarity, mean-motion resonances and/or circular orbits. These assumptions were necessary to constrain this degenerate problem to a sufficiently small orbital subspace. The priors in our analysis solely reside in the uncorrelated prior random distribution of each orbital Keplerian elements of each planet. We present our results in Section \ref{sec:Astro} and discuss them in the context of already published work in Section \ref{sec:Discuss}. In a subsequent paper we will fold these constraints on the orbits of each planet into an comprehensive dynamical analysis of this system \citep{AArronDynamocal}.\\

\section{Data reduction and methods for high-contrast astrometry with an IFS.}
\label{sec:Data}
\subsection{Observation and global instrument calibration}

HR8799 was observed with P1640 on June 14th and June 15th 2012 under excellent conditions and on October 5th 2012 under median conditions. June 14 and 15 observations comprised a total of 46 and 31 minutes of exposure time, while 165 minutes of integration time were obtained on 5 October 2012. Details of the observations and conditions and instrumental setup are thoroughly described in \citet{2013ApJ...768...24O} and we refer to that paper for further details. These data took advantage of the interferometric calibration system \citep{wbp09,2010SPIE.7736E..77P,Gautam12}. Because of their high quality, and in particular their sensitivity to the two innermost planets we only consider the June 14-15th data in the present paper. When seeking to measure the position of faint companions with respect to their host star three main sources of uncertainties arise: uncertainties associated with intrinsic instrumental calibrations (distortion, plate scale, and absolute North Orientation), uncertainties associated with the actual location of the star in the focal plane, and biases induced by the speckle suppression algorithm that is required to disentangle the exo-planetary photons from light scattered by wavefront errors.

We calibrated plate scale and absolute North using by observing the visual binary HD120476, with a grade 2 orbit, \footnote{note that while there are no reported error bars for the orbital elements of this source, uncertainties in PA and separation of this binary can be derived based on contemporaneous observations of this source with Robo AO (Riddle et al, 2014)}, from the Sixth Catalog of Orbits of Visual Binary Stars \citep{hmw01} \footnote{see http://www.usno.navy.mil/USNO/astrometry/optical-IR-prod/wds/orb for the 6 th release of this catalog} on the night of June 14th. Intrinsic distortions induced by both the PALM 3000 adaptive optics system and P1640 are by design smaller than $0.1$'' \citep{bda08,hoz11}over the small P1640 field of view. Their night to night variations have been measured to be of the same order \citet{zbo11}. We could unfortunately not obtain images of the a globular cluster in order to empirically derive a geometric distortion map that is contemporaneous with our HR8799 observations \citep{2010ApJ...725..331Y}. We mitigate this lack of an empirical distortion reference by nodding the position of our images of HD120476 over the P1640 detector and deriving a plate scale and PA offset at each location. When folding together these measurements we derived a plate-scale of $0.01948'' \pm 0.00005$'', and a rotation of the focal plane array with respect to absolute north of $108.92^{\circ} \; \pm 0.5^{\circ}$. Because the host star is hidden by the Apodized Pupil Lyot Coronagraph \citep{S05,2011ApJ...729..144S} and the four planets have relative brightness respectively of 3.2\%, 3.3\%, 2.9\% and 3.7\% of the mean speckle brightness in their vicinity, the other two uncertainties require particular scrutiny. Below we detail the methods our team developed to address potential systematic and quantify robust confidence intervals for these two sources of uncertainties. 
% As sanity check we observed three supplemental binaries which span a significant range of separations and Position Angle (HIP86974, SAO83007, HD175225) with published astrometric measurements within a year of our observing window. 

\subsection{Location of the star in the focal plane.}

\subsubsection{Relative alignment, correction of Atmospheric Differential Refraction (ADR)}
Our dataset data is composed images in which the stellar position varies as a result of Atmospheric Differential Refraction within a multi-wavelength cube and instrument tip tilt jitter between exposures. Our first step is thus to make sure that this ensemble of Point Spread Function (PSFs) is co-aligned: 
\begin{itemize}
\item due to the presence of quasi-static speckles, post-processing is needed to discriminate planets from speckles in these coronagraphic images. Precise sub-pixel image registration is a necessary condition for the algorithms discussed in Section 2.3 to yield optimal performances. 
\item measuring orbital motion is the primary goal of the present paper. Stellar location in the field of view thus ought to be estimated precisely either in each single slice of each cube or in co-added broadband images,  composed of slices that have been first been aligned in the relative sense. 
\end{itemize}

In this paper we follow the latter approach: we first compensate ADR and tip-tilt jitter by registering all realizations of the PSF in the observing sequence one to another. Once relative alignment is achieved we combined all slices in all cubes in order to estimate stellar position (see Section 2.2.2).\\

We achieve relative registration using cross-correlations between images \citep{cpb11,pcv12}. We found that this methods, solely based on the data at hand, did yield better registration on P1640 data when compared to methods based on PSF models. We first start by retrieving the scaling relationship between slices in a cube. Indeed, in an IFS data-cube the PSF of the quasi-static speckle field stretches as the wavelength increases, and this feature can be used to reveal the presence of planets below the noise floor set by wavefront errors \citep{sf02}. 

However this scaling relationship prevents us from directly applying cross-correlation based image registration algorithms between cubes at different wavelengths. Our first step towards ADR correction thus consists of stretching/squashing all slices to the same scale, usually corresponding to the reference wavelength at the spectral channel of highest throughput in P1640. While this scaling relationship is linear in theory, its behavior as a function of spectral channel can be altered by the earth's atmosphere or the instrument's dispersion. It is also preferable to  retrieve it empirically based on the data at hand, using a method that is not sensitive to stellar position. This is achieved by correlating the absolute value of the Fourier Transform of two PSFs obtained at separate wavelengths. Indeed, the absolute value of the transformed PSF in the u-v plane captures the information relative to the spatial scale of each image, and does not depend on the relative centering of the images or stellar location (which is captured by the phase in the u-v plane). Our procedure then goes as follows. The template PSF is transformed using a Matrix Fourier Transform of scaling unity (equivalent to a Fast Fourier Transform) while the PSF for which the relative scaling is sought is transformed using a Matrix Fourier Transform of scaling $\gamma$ (see details in \citet{2007OExpr..1515935S}). We then proceed to find the value of $\gamma$ which minimizes the cross-correlation of the modulus of the two transformed images. We find that, while the spatial scaling law deviates from the theoretical linear behavior over the full P1640 wavelength range ($0.98-1.75 \; \mu$m), it does not vary significantly from exposure to exposure and only needs to be updated on a night-to-night basis, or run to run, depending on observing conditions. Once this scaling law is known, we proceed to either compress or stretch all the slices in an observing sequence to our chosen reference wavelength.\\

In a second step, now that all slices are at the same scale, we carry out relative alignment by cross-correlating each slice to a reference images, chosen as the slice at our reference wavelength in the first cube of the observing sequence. We calculate the relative image alignment offsets of each channel using the sub-pixel alignment algorithm described in \cite{Guizar-Sicairos:08}. Finally we stretch/squash all cubes to their natural scale in order to obtain a new series of cubes that have been ADR and tip-tilt corrected. In this set of aligned cubes (in the relative sense) all slices in all cubes of the observing sequence are registered so that their stellar location coincides as well as possible with the stellar location in the reference slices. This corresponds to an empirical correction of both the atmospheric dispersion across each given cube, and of the tip-tilt jitter between cubes. Note that we compared the result of this relative alignment process with ADR models and found good agreement within $\pm$ 1/10 th of a pixel for this particular dataset and a variety of other P1640 observations (Nilson et al., in prep). 

However at this stage the {\bf absolute stellar location in the focal plane is unknown}. This quantity is critical to constrain the orbital motion of planets around their parent star and we next show how to estimate it using this set of empirically registered slices. \\

\subsubsection{Absolute location of the star}

\begin{figure*}[htb]
\center
\includegraphics[height = !,width=13cm]{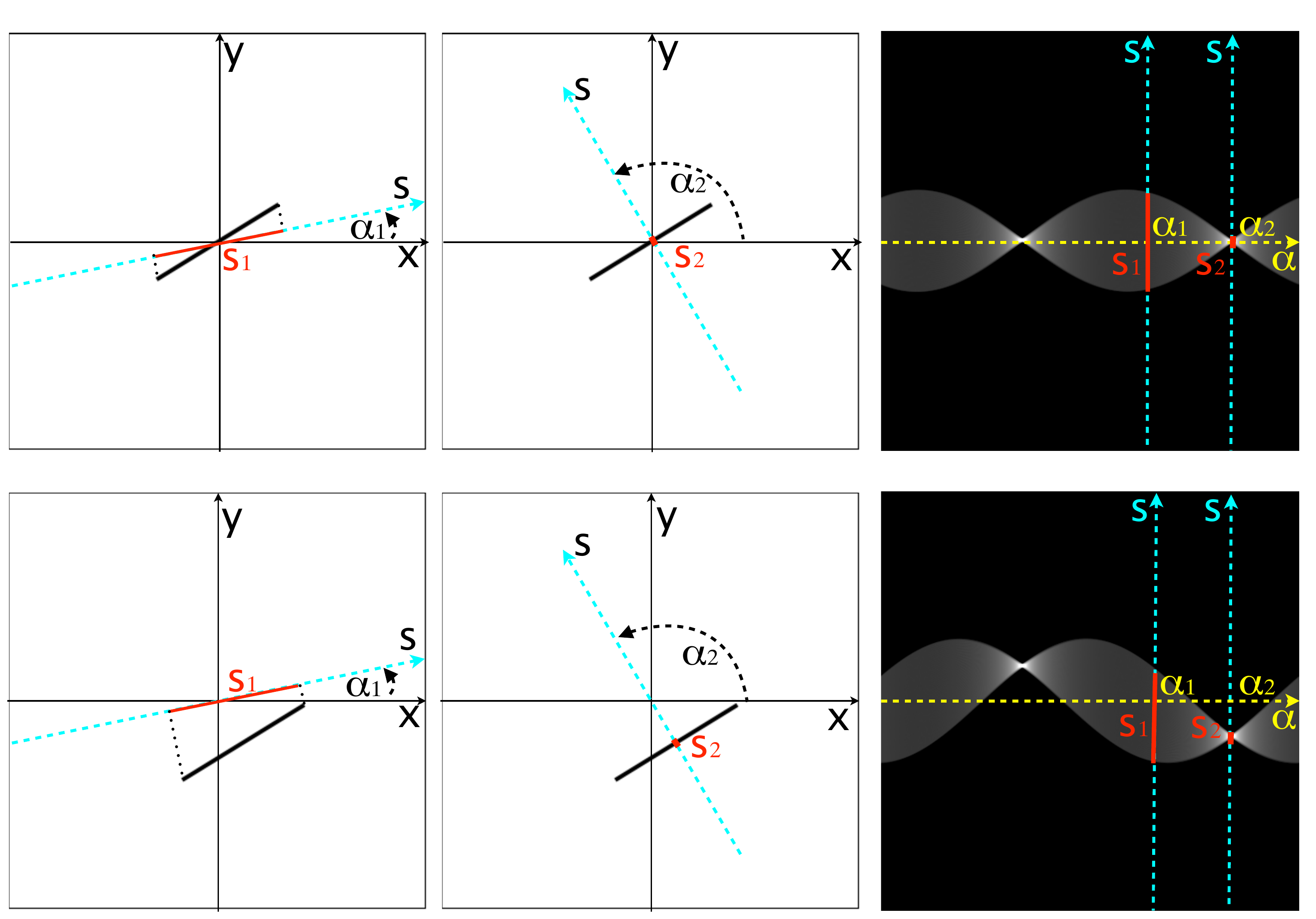}
\caption{Illustration of the radon transform of a simple line-object (black line) in the case where the line is aligned with both the coordinate center and the radon transform center $(x_0,y_0) = (x_C,y_C)$ (top three panels), and in the case where the line is shifted from the radon transform center (bottom three panels). The radon transform (right panels) represents the projection along the $s$ vector as a function of the angle $\alpha$. The left figure shows the projection for arbitrary angle $\alpha=\alpha_1$ (red line $S_1$), which can also be seen in the radon transform at angle $\alpha_1$. For the projection angle $\alpha=\alpha_2$ the projection is concentrated into a single point in the radon transform (or a very small spot if the line has a finite thickness). When the line-object is aligned with the center of the radon transform (see top three panels), the bright spots in the radon transform (at $\alpha=\alpha_2 + k \pi$, $k \in \mathbb{Z}$) are located at $s=0$. In the case where the line-object is offset from the center of the radon transform, the bright spot for projection angle $\alpha_2$ is obtained for $s\neq0$. This important property is used here to determine the precise location of the star either using the satellite spots, or broadband elongated speckles, since the radon transform becomes a superposition of these patterns.} 
\label{fig:RadonTheory} 
\end{figure*}

Current direct imaging observations rely on deep images with the host star being saturated to identify the planets. The location of the star is estimated using short exposures, combined with the introduction of a Neutral Density filter \citep{mmb08,mzk10,2013A&A...549A..52E}, or using the location of the secondary support structures \citep{Sou11}. In the presence of a coronagraph, and of a differential tip-tilt closed loop operating to ensure alignment of the star with focal plane occulting spot \citep{dho06}, there is no direct image of the star in the focal plane. This largely complicates the estimation of stellar position. To address this problem, a set of four fiducial stellar PSFs can be introduced in outer radii of coronagraphic images \citep{mlm06,so06,zoh10}, either using a pupil plane grid associated with the apodizer or by modulating the surface of the Deformable Mirror. These fiducial ``satellite spots'' then create an astrometric reference frame can be used to infer the location of the host star in occulted images. When using an astrometric grid or DM modulation we can take advantage of the broadband radial elongation of ``artificial speckles'' to infer the location of the star. The P1640 June observations of HR8799 were conducted without the pupil plane grid and without DM modulation. Here we demonstrate how the stellar location can still be derived in such images based solely on the radial elongation of ``natural speckles'' in IFS data. Our method can be applied to configurations for which fiducial astrometric spots are present, and we discuss both cases for the sake of generality. However it is important to notice that our method relies on the hypothesis that broadband speckles do point towards the stellar location. In general this is not true since ADR, or ADC residuals, does modify the stellar location across the bandpass and this hypothesis is not rigorously true. However because in the case of IFS data we reconstruct the broadband image based on narrow band slices that have already been registered one to another this effect is largely mitigated, whether or not the ADR is corrected in the instrument, and the only residual source of uncertainty corresponds to the stellar motion due the earth atmosphere within the narrow bandpass of a P1640 slice. This uncertainty is much smaller than the one arising from the method herein. Thus for the remainder of the paper we operate under  the hypothesis that broadband speckles do point towards the stellar location. For non dispersed data, the chromatic stellar motion ought first to be estimated using  ADR, or ADC residuals, models before the method described below can be carried out.

We write a co-added broadband image as $b(x,y)$ where $(x,y)$ denote focal plane coordinates. Note that the relative alignment, between spectral channels, is necessary to create such broadband cubes (e.g. we need a cube for which we have established that the star is on the same spaxel, albeit unknown, at each wavelength). In such an image both speckles and fiducial spots appear as radially elongated structures, which all point toward the stellar position, somewhere behind the focal plane mask. There are two sources of speckle (either ``natural'' or satellite spot) elongation beyond the characteristic scale of the airy dis: ADR within each narrow band and low order wavefront errors. Precisely estimating the location of a speckle in a narrow band slice would require to model both effects carefully and we did not pursue this avenue in this paper (this is a complicated exercise that will receive its own scrutiny is an upcoming paper). Instead we chose a global approach based on speckles elongation across the entire instrument bandpass. Our Radon approach does not estimate the point coordinates to which speckles point {\em at a given wavelength}, it measures the point coordinates from which each speckles move {\em across a full cube of narrow band slices}. Using P1640 data with satellite spots we compared the global approach discussed below, with more classical methods based on estimating stellar position based on spots centroid in each slices and we found good agreement within $\pm 1/10$ th pixel. \\

\begin{figure}[htb]
\center
\includegraphics[height = 19cm]{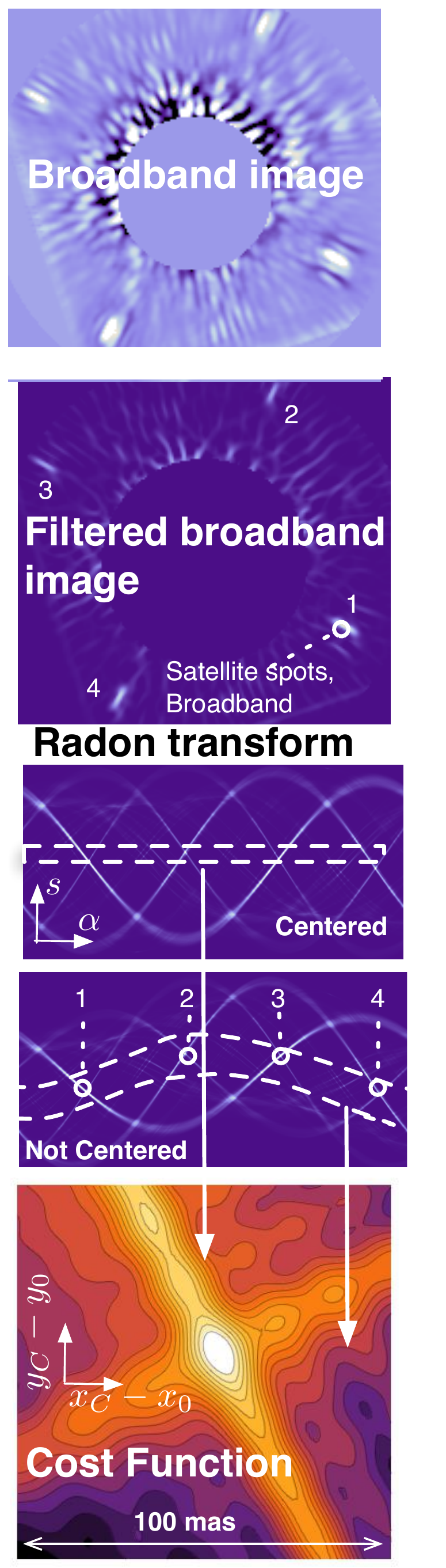}
\caption{Determination of stellar location in Radon space with fiducial spots: {\em Top:} Broadband PSF created by adding aligned hyper-spectral slices. {\em Second:} Broadband PSF propagated through a wavelet filter tuned to the characteristic scale of the four broadband fiducial lines. {\em Third:} Radon Transform of the second panel, the sinusoidal traces of each four broadband fiducial lines can be identified. When the Radon transform is calculated assuming that the center of the image coincides with stellar location then the maxima of each sinusoidal trace are located on the $s = 0$ axis. Otherwise the scatter of these maxima significantly deviates from this axis. {\em Bottom:} Cost function calculated by integrating the energy along the $s = 0$ axis in Radon space as a function image centering. The maximum of this quantity lies at the location of the star, its spread yields an estimate of the uncertainty associated with stellar position.} 
\label{fig:RadonWithSpots} 
\end{figure}

\begin{figure}[htb]
\center
\includegraphics[height = 18.5cm]{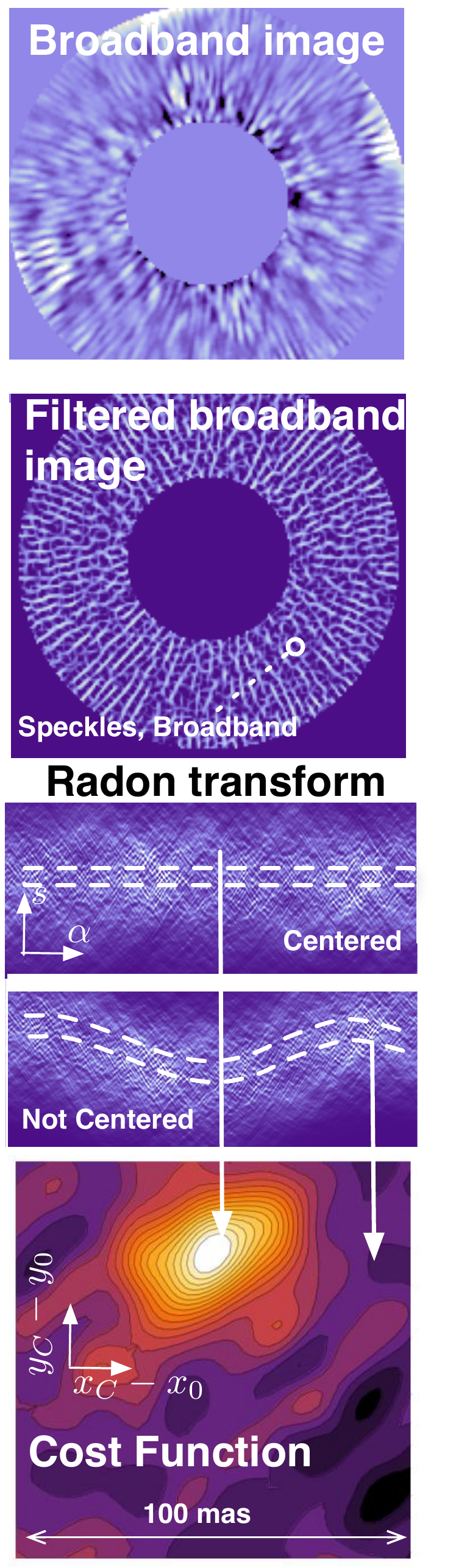}
\caption{Determination of stellar location in Radon space with ``natural speckles'': {\em Top:} Broadband PSF created by adding aligned hyper-spectral slices. {\em Second:} Broadband PSF propagated through a wavelet filter tuned to the characteristic scale of the speckles lines. {\em Third:} Radon Transform of the second panel, where the sinusoidal traces of the speckles are significantly less pronounced than in the case of fiducial spots. However when the Radon transform is calculated assuming that the center of the image coincides with stellar location then the maxima of most of the sinusoidal trace is still located on the $s = 0$ axis. Otherwise the scatter of these maxima significantly deviates from this axis. {\em Bottom:} Cost function calculated by integrating the energy along the $s = 0$ axis in Radon space as a function image centering. Because ``natural speckles'' are less pronounced than fiducial spot the uncertainty associated with stellar position is larger, however it is still well constrained.}
\label{fig:RadonWithOutSpots} 
\end{figure}

Our first step consists in using a wavelet filter to emphasize the radial structure due to speckles or satellite spots in a broadband image. The cutoff scale of this filter is tuned to the characteristic scale of the structures of interest in the broadband image: a few units of angular resolution when using fiducial spots (whose first and sometimes second airy rings are significantly brighter than the surrounding speckle floor) or a single unit of angular resolution when using ``natural speckles''. This step results in an image in which the radial structures in the broadband image have been emphasized $b_{F}(x,y)$, as shown in the top two panels of Figures \ref{fig:RadonWithSpots} and \ref{fig:RadonWithOutSpots}.\\

In order to illustrate how we derive the position of the star based on the information contained in broadband radial structures, we consider the case of an image composed of a finite number of infinitely thin radial lines of length $L$ all converging onto the point $(x_C,y_C)$:
\begin{eqnarray}\label{radialLines}
&& b_{F}(x,y) =\Pi\left(\frac{\sqrt{x^2+y^2}}{L}\right)\\ 
&& \times \sum_{k = 1}^{N_{Lines}} \delta \left[ (x-x_C) \cos \alpha_k +(y-y_C) \sin \alpha_k \right]\nonumber,
\end{eqnarray}
where $\alpha_k$ is the slope angle of each line, and $\Pi$ the top-hat function and $\delta$ the Dirac distribution. In a coordinate system centered at $(x_0,y_0)$, the Radon transform of $b_{F}(x,y)$ is given by:
\begin{eqnarray}
&& Rb_{F}(s,\alpha)[x_0,y_0] = \\
&& \int_{- \infty}^{\infty} b_{F}(t \sin \alpha + s \cos \alpha + x_0, -t \cos \alpha + s \sin \alpha + y_0) dt \nonumber,
\end{eqnarray}
The Radon transform of the image with the set of radial lines is therefore:\\
\begin{equation}
Rb_{F}(s,\alpha)[x_0,y_0] =\sum_{k = 1}^{N_{Lines}} r_k(s,\alpha)
\end{equation}
where the radon transform of a single line is the two-dimensional image shown in the right panel of Figure \ref{fig:RadonTheory} and represents the projection of the object along the $s$ vector as a function of the angle $\alpha$. In the case of an image defined as a set of radial lines described by Equation \ref{radialLines} (i.e. similar to a field of radially elongated speckles centered around the star), the radon transform becomes a superposition of these patterns for a single line at different phase angles. The radon transform is therefore mostly concentrated along a suite of bright spots corresponding to the angles orthogonal to the radial features in the image. When the center of the coordinate system underlying the Radon transform corresponds to the location of the star, i.e. $(x_0,y_0) = (x_C,y_C)$, the transform of the ensemble of lines has all bright spots located along the $s = 0$ line. Otherwise, the radon transform takes the form of a suite of ``point-like'' bright spots distributed along a trigonometric curve (see e.g. bottom three panels of Figure \ref{fig:RadonTheory}). 
In practice the radial lines are thick and of finite length (elongated speckles or astrometric spots), but the Radon transform remains very localized in the bright cores as shown in Figures \ref{fig:RadonWithSpots} and \ref{fig:RadonWithOutSpots}.

The fact that the Radon Transform maps lines onto points is exactly the property needed when seeking to use an ensemble of speckles to estimate the location of the star in a coronagraphic image. We thus estimate the location of the star by calculating the radon transform of a given broadband image over a grid of purported centers $(x_0,y_0)$, and find the location that maximizes the modulus square of the Radon transform over the $s = 0$ horizontal axis:
\begin{equation}
(x_C,y_C)=\argmax_{(x_0,y_0)} \int |Rb_{F}(0,\alpha)[x_0,y_0] |^2 d \alpha.
\end{equation}

The bottom panel of Figure \ref{fig:RadonWithSpots} shows the contour map of this metric in the case of a generic data cube with fiducial spots. Figure \ref{fig:RadonWithOutSpots} illustrates the particular case useful for this paper, where the star location is solely derived using ``natural speckles''. In this latter case, because of the noisier nature of the speckles, the constraint on stellar location is less tight than when fiducial spots are present. Moreover, when only using ``natural speckles'' there exist local maxima outside of this region of interest as shown in Figure \ref{fig:RadonWithOutSpots}, while in the case of fiducial spots the cost function is monotonically decreasing in all directions away from its maximum. However, modern coronagraphs such as P1640 provide absolute tip-tilt telemetry that is precise enough to provide a good first guess for stellar location even in the absence of satellite spots. \\

Figures \ref{fig:RadonWithSpots}-\ref{fig:RadonWithOutSpots} show that the Radon transform of broadband images provides a systematic way to estimate stellar location in IFS broadband images and constrain the uncertainties associated with it. As upcoming direct imaging instruments will use an IFS behind a coronagraph for their high-contrast surveys, this method is relevant to all of these projects. More general methods for alignment of such instruments can be found in \citet{2013ApOpt..52.3394S}, who emphasized as well the usefulness of Radon and Hough transforms for high-contrast imaging calibrations and science. The Radon method can be used both to test the instrument (e.g. calibrate potential non-common path errors between the Differential Tip Tilt channel and the final focal plane) and to bolster the astrometric precision of scientific observations. Note that while this method measures the point coordinates from which each speckles moves across a full cube of narrow band slices, residual uncertainties in the stellar locations can raise from speckle elongation within a narrow-band channel (due to ADR within each slice and low order wavefront errors). These effects broaden the thickness of the radial broadband speckles, which, in turn, transfers some energy from the core of the Randon transform to its wings. As a consequence the peak of the cost function in Figures \ref{fig:RadonWithSpots}-\ref{fig:RadonWithOutSpots} is broadened, thus impacting our uncertainties in stellar position. The comparison of both Figures \ref{fig:RadonWithSpots}-\ref{fig:RadonWithOutSpots} shows why the introduction of fiducial spots is preferable: they constrain the uncertainty associated with stellar location much more firmly. The uncertainty with fiducial spots is $\pm 0.1$ pixel while it is only $\pm 0.15$ pixels when carried out with natural speckles. However the ability to constrain stellar location with such an accuracy without fiducial spots demonstrates the advantages of this promising technique. When folding the P1640 plate scale, our analysis of the HR8799 data using the ``Radon star finder'' yields an uncertainty associated with stellar location of $\pm 0.0033''$.

\subsection{Location of the planets}

\subsubsection{Context}

In our data the four planets have respective relative brightness of 3.2\%, 3.3\%, 2.9\% and 3.7\% of the mean speckle brightness in their vicinity, aggressive post-processing is required to unravel them in the P1640 data. Contrast limitations due to quasi-static speckle were first discussed in \citet{mdn03,mld06} and a solution involving image post-processing was initially introduced by \citet{lmd07}, who devised the Locally Optimized Combination of Images algorithm (LOCI). Since then, several variations of this approach have been discussed in the literature, either to improve the contrast \citep{mmv10,cbi11,2012MNRAS.427..948A} or to minimize biases on astrophysical observables potentially introduced by the aggressive speckle reduction algorithm. This second problem has received particular scrutiny since thorough characterization of the sub-stellar objects discovered by upcoming direct imaging campaigns will lead to significant advances in our understanding of exo-planetary systems \citep{mmv10,pcv12,Sou11,2013ApJ...764..183B,2012A&A...545A.111M}. Proposed solutions rely on two concepts: modifying the least-squares cost function that was introduced in \citet{lmd07} in order to circumvent degeneracies associated with inverting an ill-posed problem, and/or calibrating the remaining biases by quantifying the effect of the speckle suppression on synthetic sources. Three recent papers \citep{2012ApJ...755L..28S,2012MNRAS.427..948A,s4} suggested that using Principal Component Analysis (PCA) to analyze direct imaging datasets can circumvent the problem of inverting a low rank matrix and also provides a framework to rigorously assess how much the astrophysical information is impacted by the speckle suppression algorithm. In \citet{2013ApJ...768...24O} we show two of these PCA based methods can accurately retrieve the spectrum of the four HR8799 planets in P1640 data. Since that paper was focused on the interpretation of the spectra we did not delve in the detail of either method, nor on their impact on astrometric estimates. In this section we describe how to conduct both photometric and astrometric characterization of faint point sources using the KLIP algorithm \citep{2012ApJ...755L..28S}.

\subsubsection{Nature of astrophysical biases}
The algorithms discussed above all rely on using a large collection of PSFs, obtained using one or several observation strategies (Angular Differential Imaging, Spectral Differential Imaging, Reference Difference Imaging) and subtracting out the quasi-static artifacts in images by fitting them in the least-squares sense to enhance the detectability of faint astrophysical signal. This process can lead to two types of biases on the photometric and astrometric estimates of the discovered sources:
\begin{itemize}
\item {\em Fitting bias}: where some of the astrophysical source signal is considered as speckle noise by the fitting algorithm (most severe when the inverse problem is ill-posed) and is mistakenly subtracted, even when there is no astrophysical signal present in the reference PSFs. 
\item {\em Cross-talk bias}: when astrophysical signal is actually present in the reference PSFs then self-subtraction can occur, leading to further biases in the information associated with these companions. 
\end{itemize}
We identified these two sources of confusion in \citet{pcv12}. In \citet{2012ApJ...755L..28S} we discussed how the fitting bias could be largely mitigated by first transforming the ensemble of reference PSFs into an orthogonal basis-set using a Karhunen Lo\`eve decomposition. However our argument relied on the assumption of an ensemble of reference images without any astrophysical signal. While this is true in the case of the HST-NICMOS data discussed in \citet{2012ApJ...755L..28S} it is not generally the case for most observations strategies. As a consequence cross-talk bias plays an important role in our P1640 data. In \citet{pcv12} we showed how modifying the least-squares cost function and forcing positivity of the fitted coefficients reduces the impact of this bias. Recently \citet{MaroisTLOCI} introduced a promising regularization strategy that is based on modeling astrophysical self-subtraction and including it as a penalty term in the least-squares speckles fitting problem. Here we present an alternative approach that builds upon the decomposition discussed in \citet{2012ApJ...755L..28S}. We illustrate its application to the case of point sources detection in Integral Field Spectrograph data, but it can be generalized to any observation strategy and extended objects. 

\begin{figure}[htb]
\center
\includegraphics[width=0.8\columnwidth]{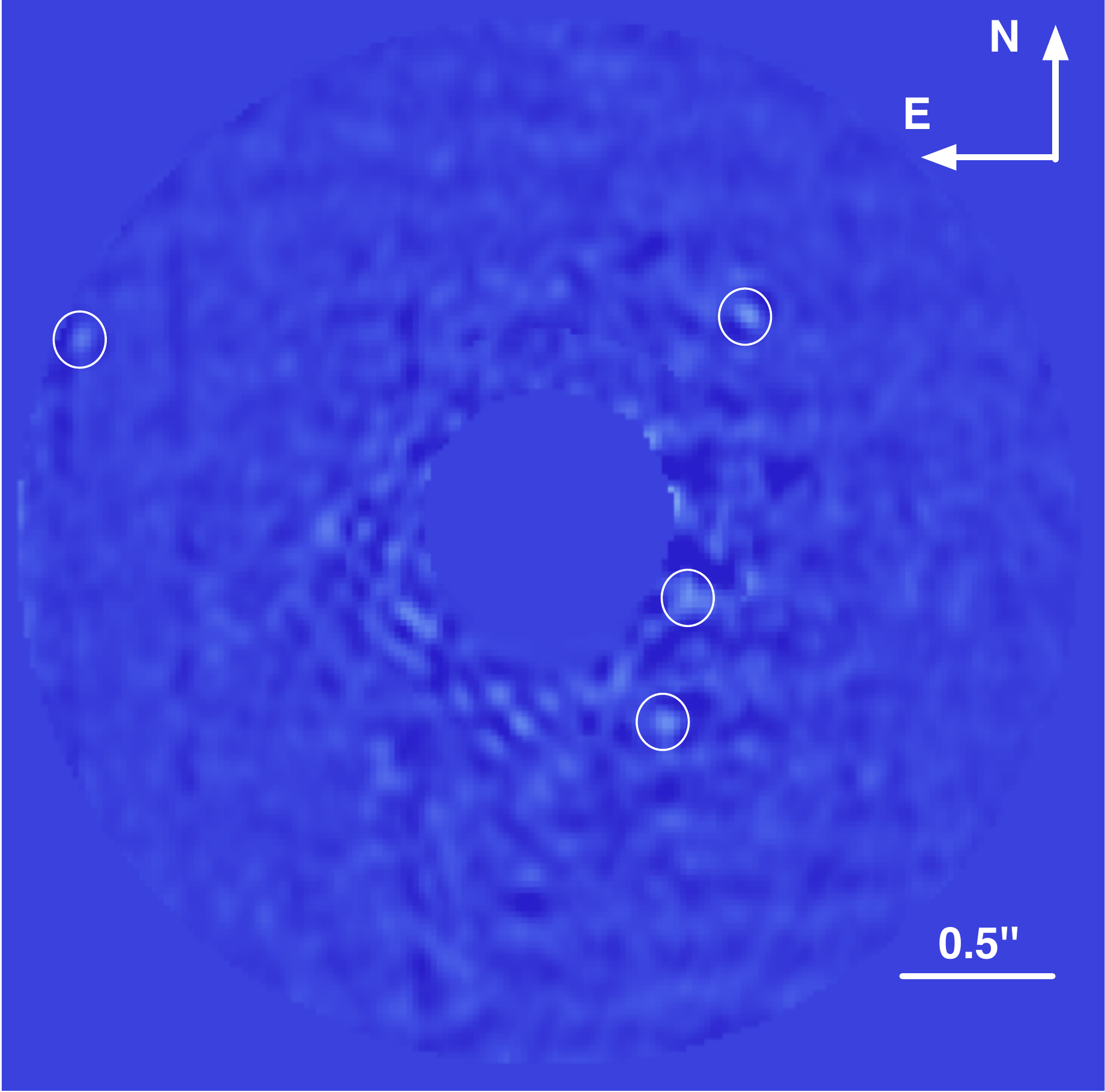}
\caption{Integrated H band image of the HR8799 system seen with the P1640 IFS. This image was created by median combining a series of reductions using the KLIP algorithm over a large set of search zone geometries, radial exclusion parameters and principal components threshold. While HR8799bcd can be detected over a very wide range of parameters HR8799e requires a fine-tuning of the azimuthal extend of the search region and the radial exclusion parameter. }
\label{fig:ImagePlanets} 
\end{figure}

\subsubsection{Reference libraries for detection and characterization}

IFS data is composed of a series of exposures obtained at times $t_p$ and at sequential wavelengths $\lambda_k$: we denote such a data-set as a collection of images $I_{t_p,\lambda_k}(x,y)$, with $p \in [1,P_{exp}]$, $k \in [1,K_{\Delta \lambda}]$, where $P_{exp}$ is the number of exposures in the observing sequence and $K_{\Delta \lambda}$ the number of spectral channels of the IFS. We assume here that all images have been centered (e.g. the relative and absolute centering described above has been carried out). In \citet{cpb11,pcv12} we showed how the detectability of faint astrophysical signal could be enhanced in an image using the LOCI algorithm with a PSF library based on rescaled images with the scaling factors resulting from the relative alignment routine. In this paper we use the KLIP algorithm \citep{2012ApJ...755L..28S} in order to detect point sources in the P1640 data. We set up the least-squares problem associated with KLIP as follows:
\begin{itemize}
\item The image of interest at $\lambda_{k_0}$ is partitioned in a series of search zones $\mathcal{S}$ centered on pixel located at $(x,y)$ of radial extent $\Delta r$ and azimuthal extent $\Delta \theta$.
\item A reference library associated to each search zone is created: for each radial location $r$ in the PSF, only a subset of references wavelengths $ \mathcal{R}_{k_0,r,N_{\delta}}$ is kept in the library. This subset of is chosen such that $|\gamma_{k_0,k} - 1| > N_{\delta} W_{\lambda_{k_0}}$ where $W_{\lambda_{k_0}}$ is the FWHM of the PSF at $\lambda_{k_0}$ and $N_{\delta}$ is a parameter tuned so that it is large enough to avoid self-subtraction (in practice $N_{\delta} \sim 1-2$ yields detection maps close to the photon noise associated with the speckles in the raw data).
\item The Principal Components, $Z_{KL}$, of the the reference library, $I_{t_p,\lambda_{k}}(x,y)$ are calculated and subtracted from the image at $\lambda_0$ according to the procedure described in \citet{2012ApJ...755L..28S}.
\end{itemize}
Carrying out this approach over a wide range of parameters $(\Delta r, \Delta \theta, N_{\delta}, K_{KLIP})$ yields deep detection maps, illustrated on Figure \ref{fig:ImagePlanets}, where an improvement in contrast of $\sim 30$ reveals the four planets orbiting HR8799 in the P1640 H band data. While very practical for detection, this approach breaks down one of the fundamental assumptions in \citet{2012ApJ...755L..28S}, namely the fact that {\em the reference library does not have any astrophysical signal located in the search zone $\mathcal{S}$}.
\begin{figure}[htb]
\center
\includegraphics[width=0.45\textwidth]{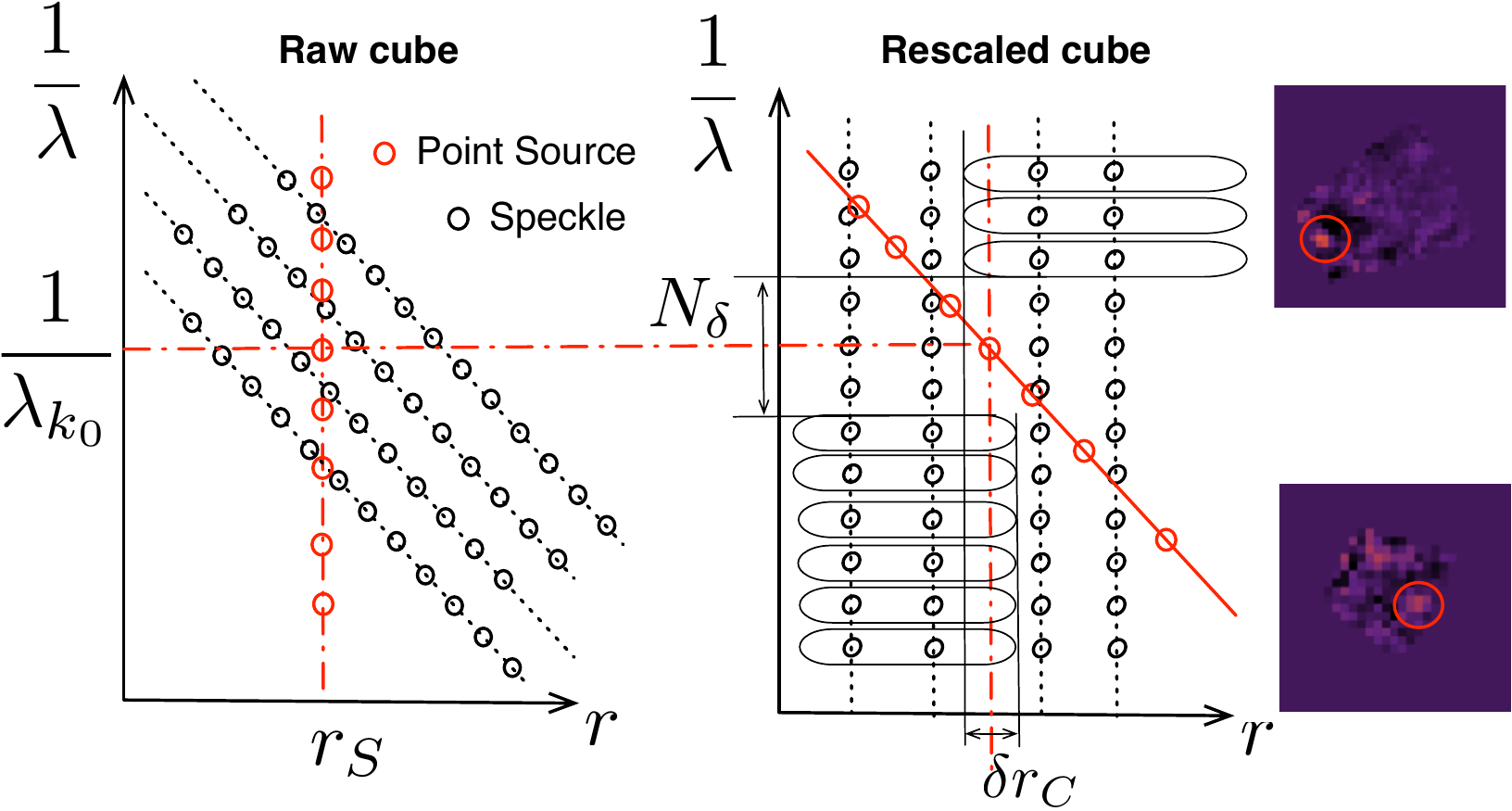}
\caption{Optimal reference PSF libraries for faint companion characterization when using IFU data without angular diversity. Once a companion has been identified and it rough location is known our goal is to focus on small regions surrounding it and build a library which contains the least amount of companion's signal possible. To do so we create a ``characterization zone'' in which the companion is located near the radial inner (or outer) edge, and only use as references re-scaled PSFs corresponding to the longer (or shorter) wavelengths. This limits the number of reference images with companion flux present to the few PSFs for which the flux at shorter (or longer) wavelength is still present between the source and the inner (or outer) edge of the characterization zone. This ensures that for each wavelength we are using a PSF library with minimal companions wavelength cross talk. We then vary the parameters $\Delta r$ and $N_{\delta}$ in order to minimize this residual cross-talk. Whether we choose the characterization zone so that the companion is near its inner or outer edge is determined by choosing the configuration which will yield the largest ``companion free'' PSF library.}
\label{fig:SlicesCartoon} 
\end{figure}
Fortunately this issue can be easily circumvented once a point source has been detected at the location $(x_p,y_p)$. One can then pose the problem so that there is little astrophysical signal from the companion candidate in the search zone by:
\begin{enumerate}
\item [(i.)] {\em either} choosing characterization zones $\mathcal{C}$ with $r \in [r_p - \delta r_C ,r_p - \delta r_C +\Delta r]$ and creating a reference library only with the wavelengths $\lambda_k \in \mathcal{R}^{+}_{k_0,r,N_{\delta}}$ such that $\gamma_{k_0,k} - 1 < - N_{\delta} W_{\lambda_{k_0}}$
\item [(ii.)] {\em or} choosing characterization zones $\mathcal{C}$ with $r \in [r_p + \delta r_C - \Delta r, r_p + \delta r_C ]$, and creating a reference library only with the wavelengths $\lambda_k \in \mathcal{R}^{-}_{k_0,r,N_{\delta}}$ such that $\gamma_{k_0,k} - 1 > N_{\delta} W_{\lambda_{k_0}}$.
\end{enumerate}
Figure \ref{fig:SlicesCartoon} illustrates these two configurations, for which there is no signal from the detected faint companion in the reference library. $\delta r_C$ denotes the radial offset between the position of the companion and the edge of the characterization (in the direction of cross talk). $N_{\delta}$ and $\delta r_C$ are the {\em only reduction parameter} that can yield cross-talk bias. They ought to be chosen carefully within the bounds $ W_{\lambda_{k_0}} < \delta r_C < N_{\delta} W_{\lambda_{k_0}}$. Aside from having to choose yet another reduction parameter, which can actually be done relatively easily as discussed below, the main drawback of this method is that it significantly reduces the number of references available for characterization when compared to the detection algorithm. In the absence of field rotation, we mitigate this effect by choosing strategy (i.) described above for blue wavelengths $\lambda_{k_0} <\lambda_{ K_{\Delta \lambda /2}}$ and strategy (ii.) for red wavelength $\lambda_{k_0} >\lambda_{ K_{\Delta \lambda /2}}$. When using all the cubes from the observing sequence in the reference library we find that in practice this approach yields levels of speckle suppression comparable to the detection pipeline, except for the few wavelengths near the middle of the spectral band-pass. In the case of P1640, speckle suppression in these few channels are of lesser importance since the middle of P1640's spectral band-pass is located in the atmospheric water band, for which telluric absorption cannot be precisely calibrated given the moderate resolution of the spectrograph. For future instruments such as GPI or SPHERE this will be mitigated by the ADI observing mode, which will enable to combine the radial offset of characterization $\mathcal{C}$ zones described above with an azimuthal offset, and thus significantly enlarge the characterization ``companion-free'' reference library.

\subsubsection{Forward modeling: principles}
Once an adequate PSF library has been set up, with a minimal amount of companion signal in the reference PSF characterization zone, the forward modeling approach suggested in \citet{2012ApJ...755L..28S} can be carried out. For simplicity we use the notations in \citet{2012ApJ...755L..28S}: the target image at wavelength $\lambda_0$ and exposure $t_{0}$ is written as $T(x,y) = I_{t_0,\lambda_{k}}(x,y)$, the ensemble of references $R_{q}(x,y)= I_{t_p,\lambda_{k}}(x,y)$ with $\lambda_k \in \mathcal{R}_{k_0,r,N_{\delta}}, \; p \in [1,P_{exp}]$. The set of principal components of this library is $Z^{KL}_q (x,y)$. The reduced image is then:
\begin{equation}
F(x,y) = T(x,y) - \sum_{q= 1}^{K_{klip}} <T,Z^{KL}_q>_{\mathcal{C}} Z^{KL}_q(x,y).
\end{equation}
We assume a known model $S(x,y)$ of a point source PSF at wavelength $\lambda_{k_0}$ with a normalized flux. Under the assumption that the noise in the reduced image $F(x,y)$ is Gaussian and of zero mean, a least squares estimator yields unbiased values for the brightness and location of a point source. In order to accommodate for the fact that this assumption might not be true over the entire characterization zone $\mathcal{C}$, we can write this cost function over a fitting region $\mathcal{F} \in \mathcal{C}$. We find that for our HR8799 P1640 data, a fitting zone that spans the entire characterization zone yields unbiased single channel astrometry and photometry, (Figure \ref{fig:Biases_At_Point5mas}). However this might not be the case for other instruments, or for fainter sources in P1640 data. In the general case of distinct fitting and characterization zones, we solve for the location of the point source $(\tilde{x}_{s},\tilde{y}_{s})$ and its flux $\tilde{f}_{s}$ by minimizing the least forward modeling cost function described in \citet{2012ApJ...755L..28S}:
\begin{eqnarray}
& & (\tilde{x}_{s},\tilde{y}_{s},\tilde{f}_{s})=\argmin_{(x_s,y_s,f_s)} \\
&& \Bigg. \Bigg| \Bigg| F - f_S \left(S-\sum_{q= 1}^{K_{klip}} <S(\cdot-x_S,\cdot-y_S),Z^{KL}_q>_{\mathcal{C}} Z^{KL}_q \right)\Bigg| \Bigg|^2_{\mathcal{F}} \nonumber,
\end{eqnarray}
and where the $\cdot$ represents the dummy integration variable for the inner product. 
Since this is a quadratic cost function, the point source coordinates can be estimated using a matched filter approach, as the location of maximum of the cross correlation between the reduced image and the PSF model propagated though the PCA filter:
\begin{equation}
(\tilde{x}_{s},\tilde{y}_{s})= \argmax_{(x_s,y_s)}\{ C_{(x_s,y_s)}(F,S,K_{klip}) \} \label{eq:maxCorr}.
\end{equation}
with
\begin{eqnarray}
&& C_{(x_s,y_s)}(F,S,K_{klip}) = <F,S(x-x_S,y-y_S)>_{\mathcal{F}}- \nonumber \\
&& \sum_{q= 1}^{K_{klip}} <S(x-x_S,y-y_S),Z^{KL}_q>_{\mathcal{C}} <F,Z^{KL}_q>_{\mathcal{F}} \label{eq:CorrExp},
\end{eqnarray}
where we have indicated the integration variables $x$ and $y$ in the terms where the explicit search variables $x_S$ and $y_S$ appear.
Once the true location of the point source $(\tilde{x}_{s},\tilde{y}_{s})$ has been estimated, its flux is then given by ratio of the reduced image-model PSF correlation and the estimate of the flux loss due to the fitting bias:
\begin{eqnarray}
&& f_s = \label{eq:FLuxEst} \\
&& \frac{C_{(\tilde{x}_{s},\tilde{y}_{s})}(F,S,K_{klip})}{|| S-\sum_{q= 1}^{K_{klip}} <S(\tilde{x}_{s}-x_S,\tilde{y}_{s}-y_S),Z^{KL}_q>_{\mathcal{C}} Z^{KL}_q ||^2_{\mathcal{F}}} \nonumber
\end{eqnarray}
Equations \ref{eq:maxCorr} to \ref{eq:FLuxEst} describe how to take advantage of the KLIP algorithm to derive the focal plane location and the brightness of a faint point source hidden in under speckles in high-contrast imaging data. The major conditions for these equations to be valid (and yield astrophysical estimates that are un-biased) are that no companion signal is present in the portion of the image used for characterization, and that the residual noise in the fitting zone is Gaussian of zero mean. We earlier showed how to build a PSF library using IFS data, which {\em almost} satisfies the former. The ability to reach gaussian noise after ADI subtractions was demonstrated by \citet{2008ApJ...673..647M}; this result remains true for LOCI- or KLIP-based subtractions, and can be achieved by choosing adequate geometries for characterization zones and tuning $K_{KLip}$ (e.g. using a thorough parameter search to minimize residual speckle noise as in \citet{Sou11}). 
In the following section, we test the accuracy of the approach described by Equations \ref{eq:maxCorr} to \ref{eq:FLuxEst} using a synthetic dataset that strictly satisfies this condition. Then, in Section \ref{sec:Astro} we illustrate how to practically derive astrophysical observables and their uncertainties using our HR8799 dataset, for which the ``companion free PSF library'' condition is not strictly true. In particular we discuss how to mitigate the residual biases that are introduced in this realistic case. 
\subsubsection{Forward modeling: results}
\begin{figure*}[!t]
\center
\begin{tabular}{cc}
{\bf \large Single wavelength baises at $0.5$'' }& {\bf \large Single wavelength baises at $1$'' }\\
\includegraphics[height = 21cm,width=8cm]{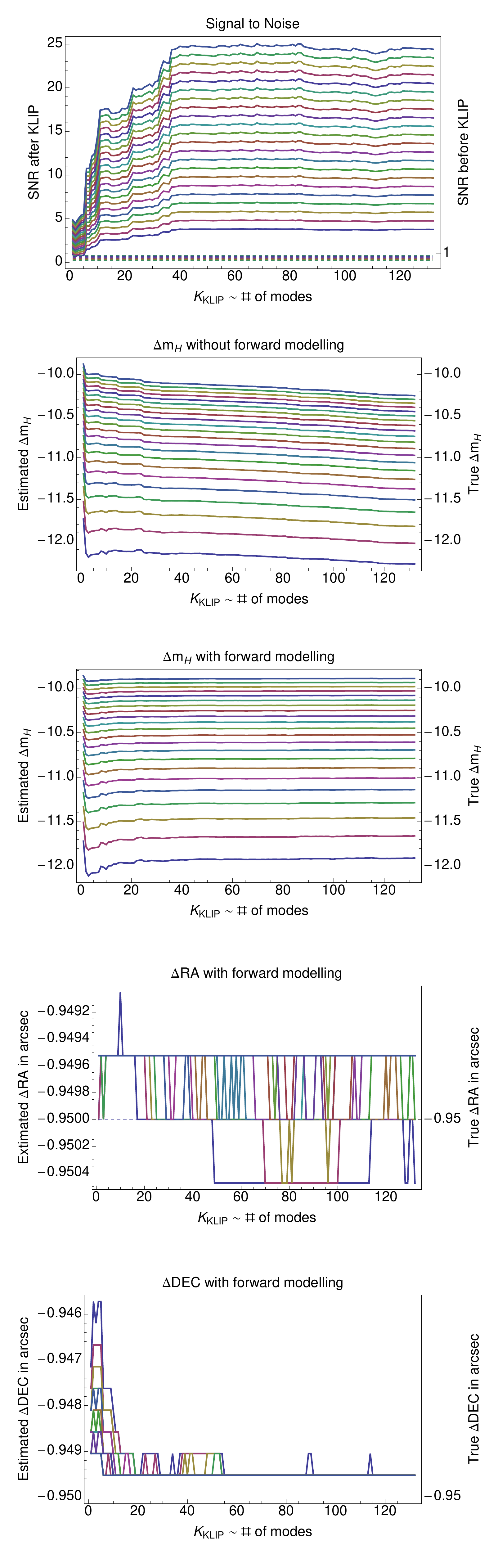}
&\includegraphics[height = 21cm,width=8cm]{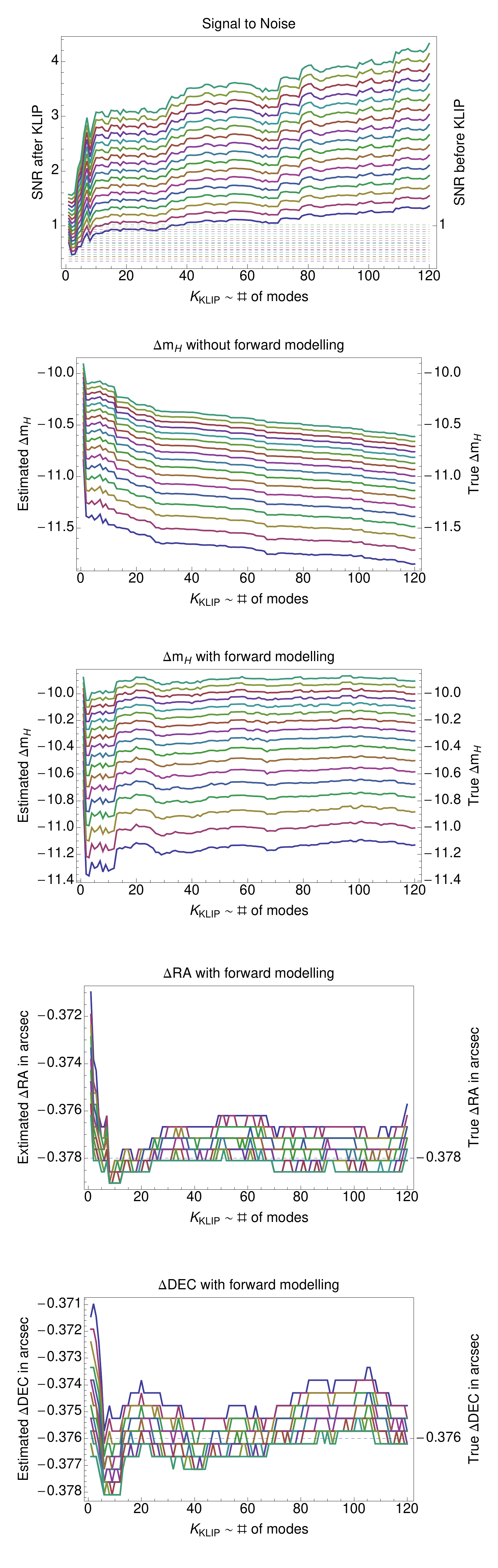}
\end{tabular}
\caption{{\bf Left: Single wavelength photometric and astrometric biases at $0.5''$}: we injected a series of synthetic point source of various brightness in the $1.3 \; \mu m$ P1640 spectral channel. For this test we {\em did not} inject any companion flux in other wavelength slices of the IFS data-cube: this represents the ideal of a reference cube that is $100 \%$ companion free. We consider the case of the fitting zone $\mathcal{F}$ which overlaps with the characterization zone $\mathcal{C}$. Without forward modeling the photometry of the reduced data is underestimated because of self-subtraction: as the number of modes increases, the SNR of the point source increases however its estimated flux becomes more and more biased. On the other hand photometry is preserved using the forward modeling to take into account of this algorithmic flux depletion. Moreover, once a sufficient number of modes has been taken into account, the forward modeling yields very precise estimates of the synthetic sources, e.g position in the focal plane. {\bf Right: Single wavelength photometric and astrometric biases at $1''$}: same exercise as on the left except that now the synthetic point source is detected with a higher statistical significance. Note the spectrophotometric and astrometric biases are further reduced in this case.}
\label{fig:Biases_At_Point5mas} 
\end{figure*}
Figures \ref{fig:Biases_At_Point5mas} illustrates the result of this approach on a synthetic companion injected in P1640 data. The purpose of these simulations is to assess the accuracy of the forward modeling under the ``companion free PSF library'' assumption, and our test data is thus built as such (e.g. using a synthetic companion that is only present in one spectral channel). The findings of this numerical experiment can be summarized as follows:
\begin{itemize}
\item {\em Photometric bias:} In our test cases, forward modeling with KLIP circumvents the algorithmic flux depletion due to fitting companion's flux with speckles discussed in \citet{pcv12}. It yields photometric estimates with a bias smaller than $0.02$ magnitudes. 
\item {\em Photometric uncertainty:} For a source with SNR $\sim 3$, the photometric uncertainty due to the algorithm (photometric scatter across number of modes $K_{KLIP}$ after forward modeling) is $\sim 0.05$ magnitudes. For sources with SNR $\sim 20$ the photometric uncertainty due to the algorithm is $< 0.01$ magnitudes. Note that these numbers correspond to the part of the parameter space for which the point source detectability is somewhat independent on the number of modes used in the PSF subtraction ($K_{KLIP} >40$).
\item {\em Astrometric bias:} In our test cases, forward modeling with KLIP yields astrometric estimate a bias smaller than $0.0005''$, corresponding to $\sim 1/40$ pixel.
\item {\em Astrometric uncertainty:} Once the number of modes is sufficient then the astrometric uncertainty due to the algorithm (astrometric scatter across number of modes after forward modeling) is $0.001''$ for sources with SNR $\sim 3$ and $0.0005''$ ($\sim 1/40$ pixel) for sources with SNR $\sim 20$.
\end{itemize}
Note that these results were obtained when using the a fitting region $\mathcal{F}$ that is exactly the size of the characterization zone $\mathcal{C}$: this choice was driven by the fact that the residual noise in the characterization zone chosen for our test data was indeed Gaussian of zero mean. When it is not the case, then one can either use a subset of the characterization region as the fitting region or simply change the geometry of $\mathcal{C}$ in an attempt to improve the residual noise statistics. We tested both approaches and did not observe any fundamental differences, and we decided that in practice we would solely use the second approach in our P1640 pipeline for the sake of simplicity.\\

Naturally, when using real data, the ideal performances above will be severely impacted when the various assumptions underlying the simulations in Figure \ref{fig:Biases_At_Point5mas} do break down. Namely: the PSF model used for the forward modeling will not strictly be equal to the actual companion's PSF, the residual speckles after KLIP might not be Gaussian of zero mean and the PSF library will not be completely ``companion free''. In particular, the absence of field rotation of P1640 results into PSF libraries that are solely based on wavelength diversity. The P1640 chromatic lever arm is actually not large enough to ensure that the ``companion free references'' condition is strictly enforced, in spite of our careful selection of geometries and PSFs libraries described in Figure \ref{fig:SlicesCartoon}. We present in Section \ref{sec:Astro} our methodology to derive spectro-photometric and astrometric estimates even when these assumptions are only loosely met and illustrate our methodology in the case of HR8799bcde. 

\section{Astrometry and orbital motion of HR8799bcde}
\label{sec:Astro}

\subsection{Position of planet in detector coordinates}
\begin{figure*}[htb]
\center
\begin{tabular}{cc}
{\bf \large HR8799b}&{\bf \large HR8799c}\\
\includegraphics[width=8.5cm]{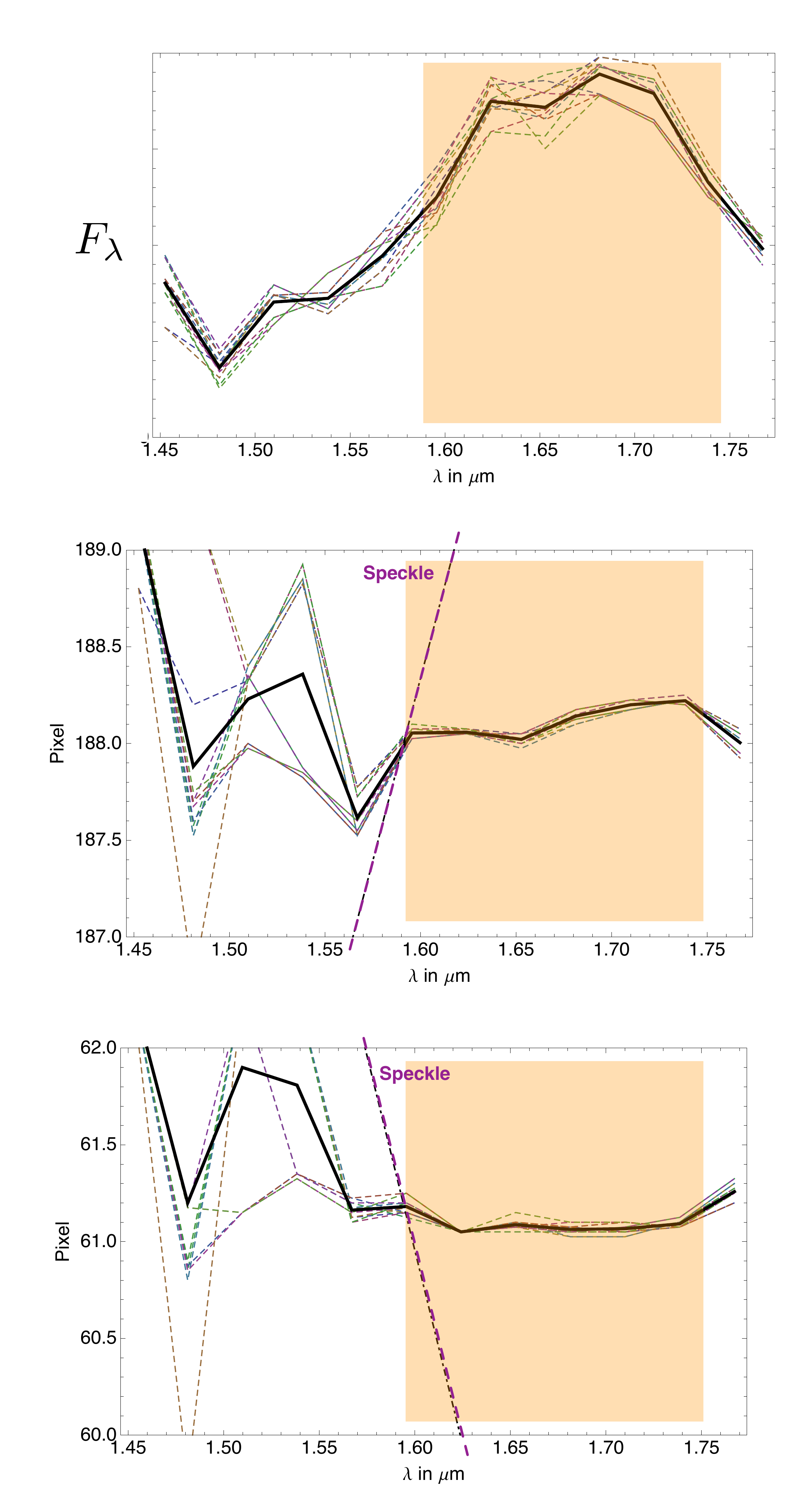}
&\includegraphics[width=8.5cm]{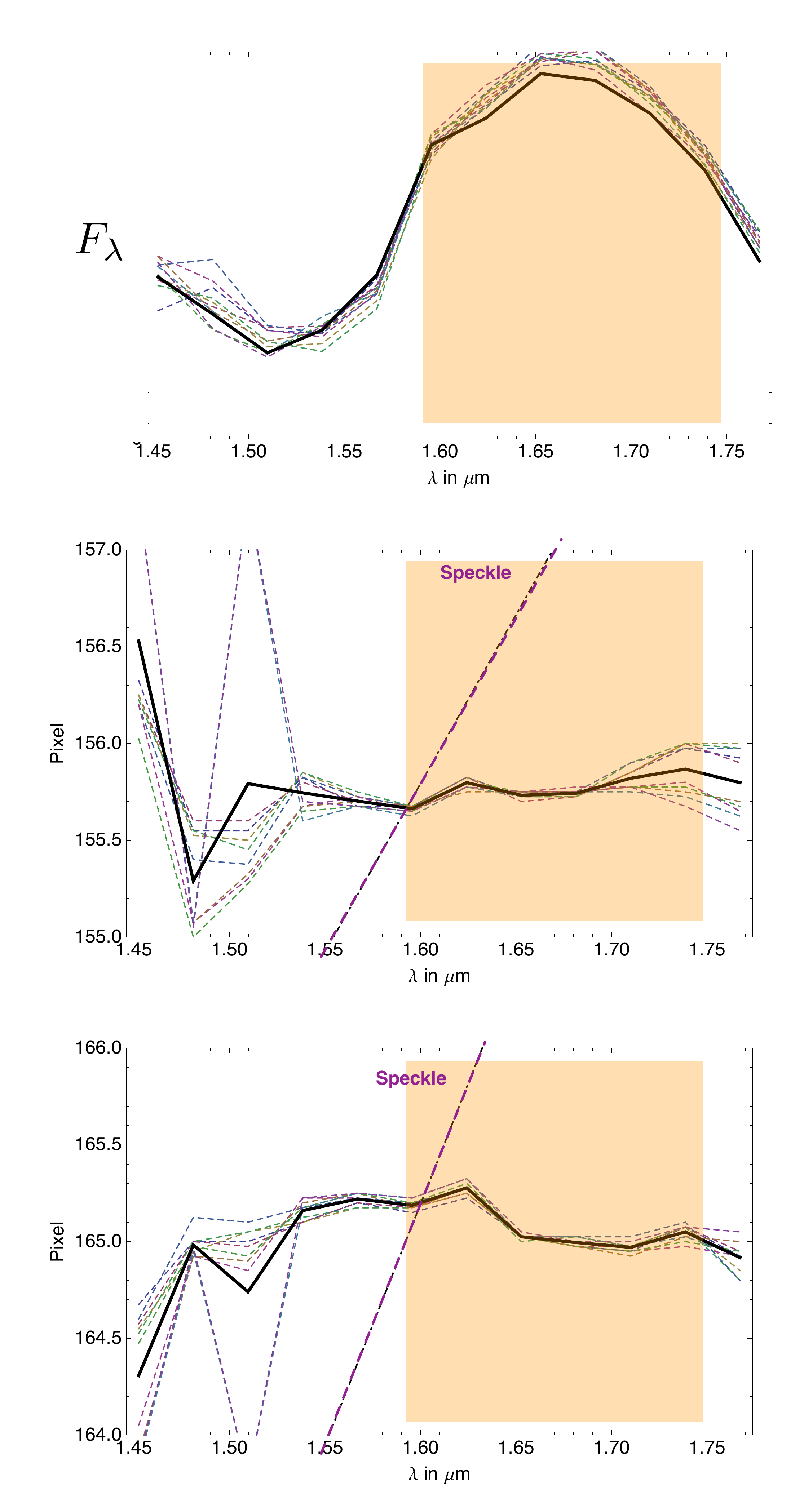}\\
\end{tabular}
\caption{{\bf Left: Spectro-photometry and astrometry of HR8799b in H band}. {\em Top}: H band spectrum of HR8799b seen with P1640: each line corresponds to a given data reduction with fixed zone geometry, exclusion parameters and $K_{KLIP}$. These spectra were selected according to the criteria described in the text and are anchored in the regime for which the astrophysical estimates cannot be biased by residual speckle or by overly aggressive PSF subtraction. The shaded band corresponds to the channels for which the signal is more prominent. {\em Bottom two panel}: $(X,Y)$ coordinates of HR8799b in detector space as a function of spectral channel estimated using KLIP forward modeling. The location of the point source in the shaded region is fixed and does not follow the radial trace of potential residual speckle at this location (dashed dot line). {\bf Right: Spectro-photometry and astrometry of HR8799c in H band}: same as left panel in the case of HR8799c.}
\label{fig:HR8799bcAstrometry} 
\end{figure*}

\begin{figure*}[htb]
\center
\begin{tabular}{cc}
{\bf \large HR8799d}&{\bf \large HR8799e}\\
\includegraphics[width=8.5cm]{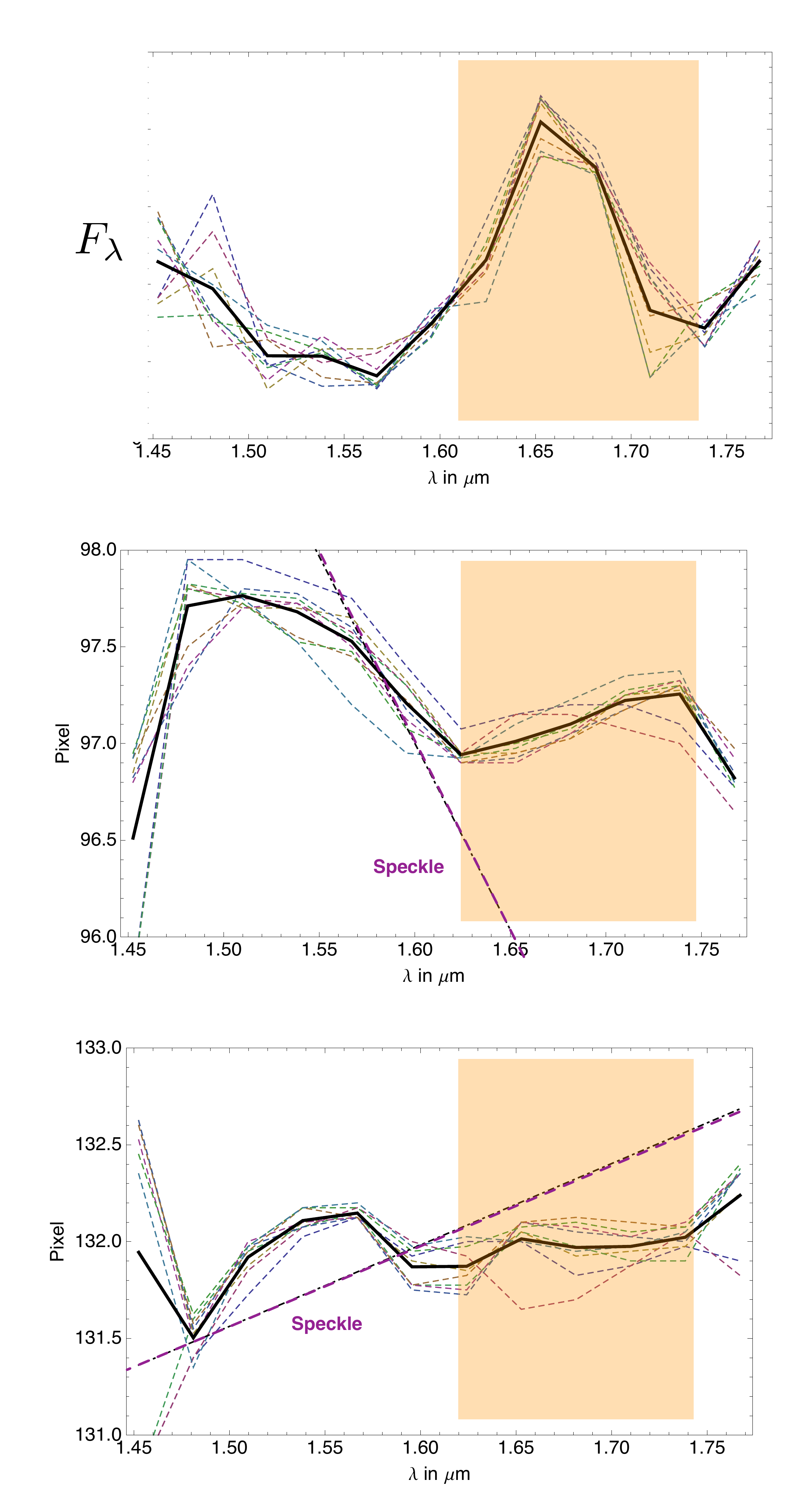}
&\includegraphics[width=8.5cm]{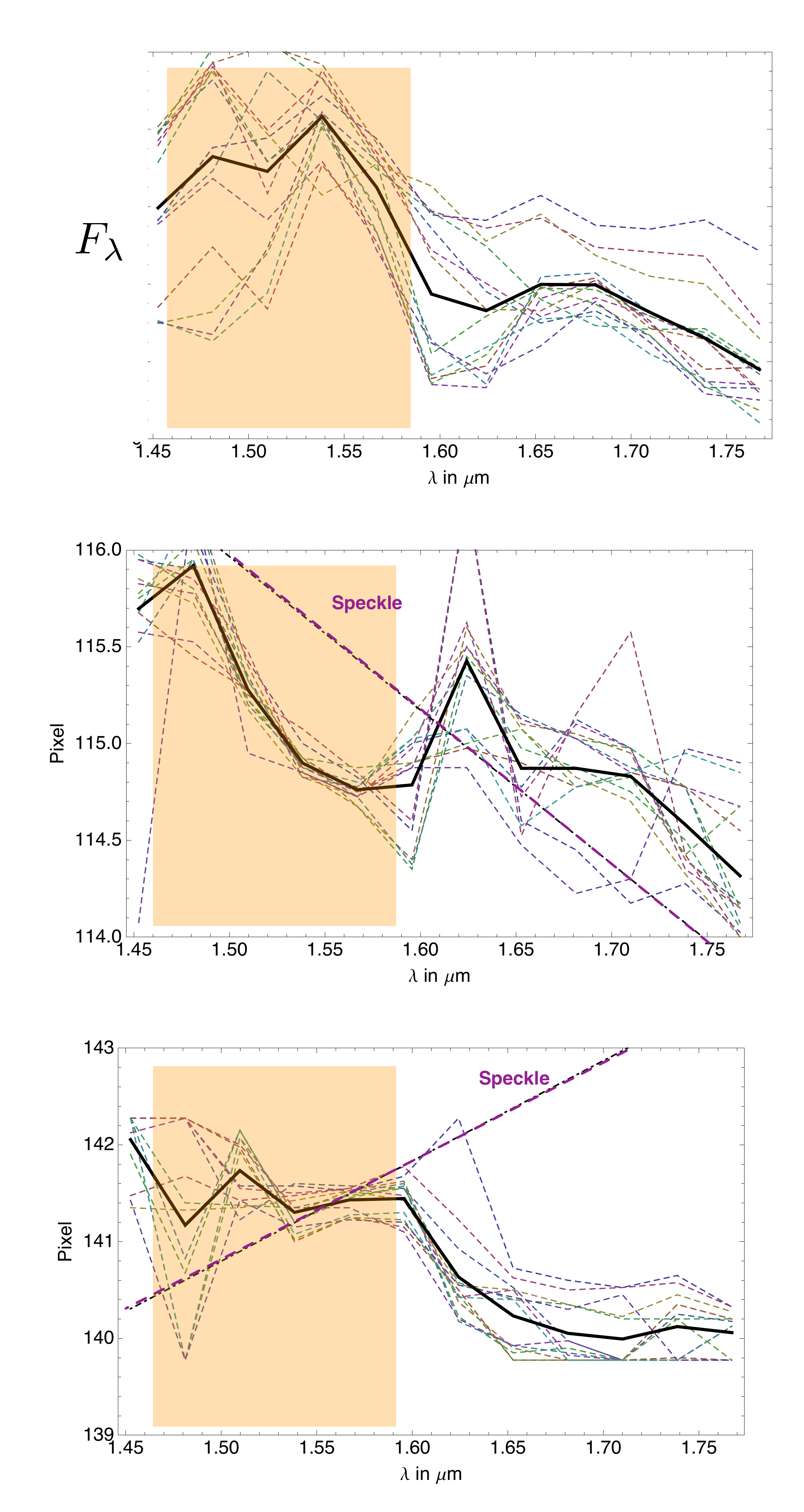}\\
\end{tabular}
\caption{{\bf Left: Spectro-photometry and astrometry of HR8799d in H band}. {\em Top}: H band spectrum of HR8799d seen with P1640: each line corresponds to a given data reduction with fixed zone geometry, exclusion parameters and $K_{KLIP}$. These spectra were selected according to the criteria described in the text and are anchored in the regime for which the astrophysical estimates cannot be biased neither by residual speckle nor by overly aggressive PSF subtraction. The shaded band corresponds to the channels for which the signal is more prominent. {\em Bottom two panel}: $(X,Y)$ coordinates of HR8799b in detector space as a function of spectral channel estimated using KLIP forward modeling. Because the statistical significance of the detection is lesser than in the case of HR8799bc the scatter on the location of HR8799d, in the shaded region, has increased. However its overall trend does not follow the radial trace of potential residual speckle at this location (dashed dot line) which ensures that we are indeed detecting an actual astrophysical source, albeit with a larger uncertainty in its astrometric location. {\bf Right: Spectro-photometry and astrometry of HR8799e in H band}: same as left panel in the case of HR8799e. The detection is now close to marginal and the uncertainty in the source location become quite large. However it is the only point source in this area of the focal plane whose astrometric signature as a function of wavelength does not follow the radial trace of a residual speckle. }
\label{fig:HR8799deAstrometry} 
\end{figure*}
In a recent paper \citet{Sou11} illustrated using HST-NICMOS data how overly aggressive PSF subtraction could substantially bias astrometric estimates. We showed on Figure~\ref{fig:Biases_At_Point5mas} that in the case of a companion free PSF library and a reduced image with zero mean Gaussian residual noise, astrometric biases were mitigated when using the KLIP algorithm. In reality, in spite of all the efforts described above, there is no guarantee that these assumptions will strictly hold. We thus resort to exploring the algorithmic parameter space in order to identify the configurations for which these assumptions are met as well as possible. Our estimates for spectro-photometry and astrometry of the faint source correspond to the average of these observables over the ensemble of ``well behaved'' reductions. Note however that this is not the same ``blind'' parameter search described in \citet{Sou11}, since we have identified the main assumptions whose validity we are seeking to test or a given dataset. We thus vary:
\begin{itemize}
\item The geometry of the characterization zone $\mathcal{C}$ in order to test the residual speckles’ statistics (e.g whether or not they are Gaussian of zero mean). Note that the size of the fitting region $\mathcal{F}$ could also be varied as well, however we find that for P1640 data changing the size of the geometry of the characterization zone is sufficient. 
\item The parameters $(N_{\delta},\delta r_C)$, in order to test the validity of the ``companion free'' assumption. Note that varying $N_{\delta}$ also impacts the efficiency of the speckle suppression (and thus the statistics of the residual speckles) since having more correlated PSFs (e.g. corresponding to a wavelength as close as possible to science wavelength) generally yields less noisy post-subtraction residuals. 
\item The number of principal components utilized for the subtraction, $K_{KLIP}$ which, will impact the statistics of the residual speckles and also provides a very insightful diagnostics tool. 
\end{itemize}
In general, for a given characterization zone geometry and given $(N_{\delta},\delta r_C)$ the behavior of spectro-photometry and astrometry as a function of $K_{KLIP}$ can be divided in three regimes:
\begin{enumerate}
\item[(a)] When $K_{KLIP}$ is small, the estimated companion's location varies both across wavelength and values of $K_{KLIP}$, which means that the PSF subtraction is not aggressive enough and residual speckles are biasing the astrometric estimate. 
\item[(b)] When $K_{KLIP}$ is large then the companion flux substantially changes with $K_{KLIP}$. This means that some of the companion's flux in nearby spectral channels of interest is actually included in the principal components. This occurs because the PSF library is not completely ``companion free''. While the small contribution of the companion to the PSF library is not captured by the first modes of the Karhunen-Lo\`eve decomposition, its influence on the estimated spectrum starts to be more prominent when $K_{KLIP}$ is large enough. 
\item[(c)] When $K_{KLIP}$ is in an intermediate regime then neither the astrometry nor the spectro-photometry vary with $K_{KLIP}$: this is the part of the parameter space that is useful to infer astrophysical estimates. 
\end{enumerate}
The boundaries between these regimes is a function of the geometry of the characterization zone, of $(N_{\delta},\delta r_C)$, and of course of the statistical significance of the companion's detection. We thus execute a parameter space search over 16 different combinations of $(N_{\delta},\delta r_C,\Delta r, \Delta \theta)$ near the detected location of the companion, and vary the number of eigenmodes in the PSF subtraction from 1 to 130. This parameter search results in $\sim$ 2000 spectra and detector coordinates for each one of the HR8799 planets. From these $\sim$ 2000 spectra, we discard any values that are either in regime (a) for which the astrometric position of the companion is not consistent between wavelength channels or (b) in which a sharp flux drop is detected with a small change in the number of modes. Finally we further trim the subset (c) by only keeping the spectra associated with images that exhibit a {\em local} SNR $>3$. Note that for this paper focused on astrometry we only consider the H band data, since P1640 detections in that bandpass present the highest statistical significance for HR8799bcde.

Note that part of this parameter search could be alleviated by using the t-LOCI technique developed by \citet{MaroisTLOCI}, which adds a penalty term to the least-squares inversion problem to take into account the predicted location of companion's flux in the reference images. However this requires the use of a template spectrum in order to operate optimally: while it presents a substantial gain by reducing the algorithmic parameter search, it is hampered by the need to test a suite of hypothesis regarding the nature of the atmosphere of the astrophysical source. Here we chose to limit ourselves to methods that {\em do not make any assumption on the properties of the detected companion} and to discuss the trade-offs associated with of spectro-photometry and astrometry in IFS data. While the boundaries of between the regimes (a), (b) and (c) can seem ad hoc, the transition between each regime can be easily quantified and the above procedure can be automated. We have not done so in the early stages of Project 1640, in order to gain insights regarding the synergy between various observing scenarios PCA based reduction methods. Our team has now successfully extracted spectra and astrometry of a handful of faint sources four orders of magnitude fainter than their host star \citep{2013ApJ...768...24O,2013arXiv1309.3372H}, and we will soon proceed to an automation of this process.

Figure \ref{fig:HR8799bcAstrometry} shows a sample of the ensemble of H band spectra and astrometry for HR8799bc that belongs to the regime (c) of ``well behaved'' astrophysical observables. Since they are detected at SNR $>10$ the residual spectro-photometric scatter in this regime in this regime is of the order of $10 \%$ which is more accurate than the precision reported by the d-LOCI algorithm at these levels of contrast \cite{pcv12}. Moreover, for the channels with the most signal, the astrometric scatter is below $0.2$ pixels ($\sim 0.004''$ for P1640), which is quite remarkable given the flux ratios between planets and raw speckles. We conservatively derive our ``detector based'' astrometric error bars as the standard deviation of all estimates in the regime c), over all wavelength in the shaded regions. The spectro-photometric error bars in \citet{2013arXiv1309.3372H,2013ApJ...768...24O} are also derived as the standard deviation of the all the spectra in ensemble (c). The case of HR8799de illustrates how to estimate uncertainties when the astrometric scatter becomes larger due to lesser signal to noise in the detection. Figure \ref{fig:HR8799deAstrometry} displays the ensemble (c) for the two inner planets in the HR8799 system: the variation of astrophysical observables across wavelength and $(N_{\delta},\delta r_C,\Delta r, \Delta \theta, K_{KLIP})$ is significantly larger in this ``well behaved regime'' than in the case of HR8799bc. This can be easily explained by the necessity to use small values of $N_{\delta}$ in order to detect these objects in our P1640 data, which yields a reference library that is more contaminated by residual planet flux at adjacent wavelength. As a final sanity check we overlay on top of the astrometric estimates the radial trace of a putative residual speckle at this location of the focal plane. Using this information we can establish that these are only point sources in their respective neighborhoods of the focal plane whose astrometric signature as a function of wavelength does not follow the radial trace of a residual speckle. We derive error bars on the spectro-photometry and astrometry in a similar fashion as we did for HR8799bc.
\subsection{Astrometry}
\begin{figure*}[t!]
\center
\includegraphics[width=8 cm]{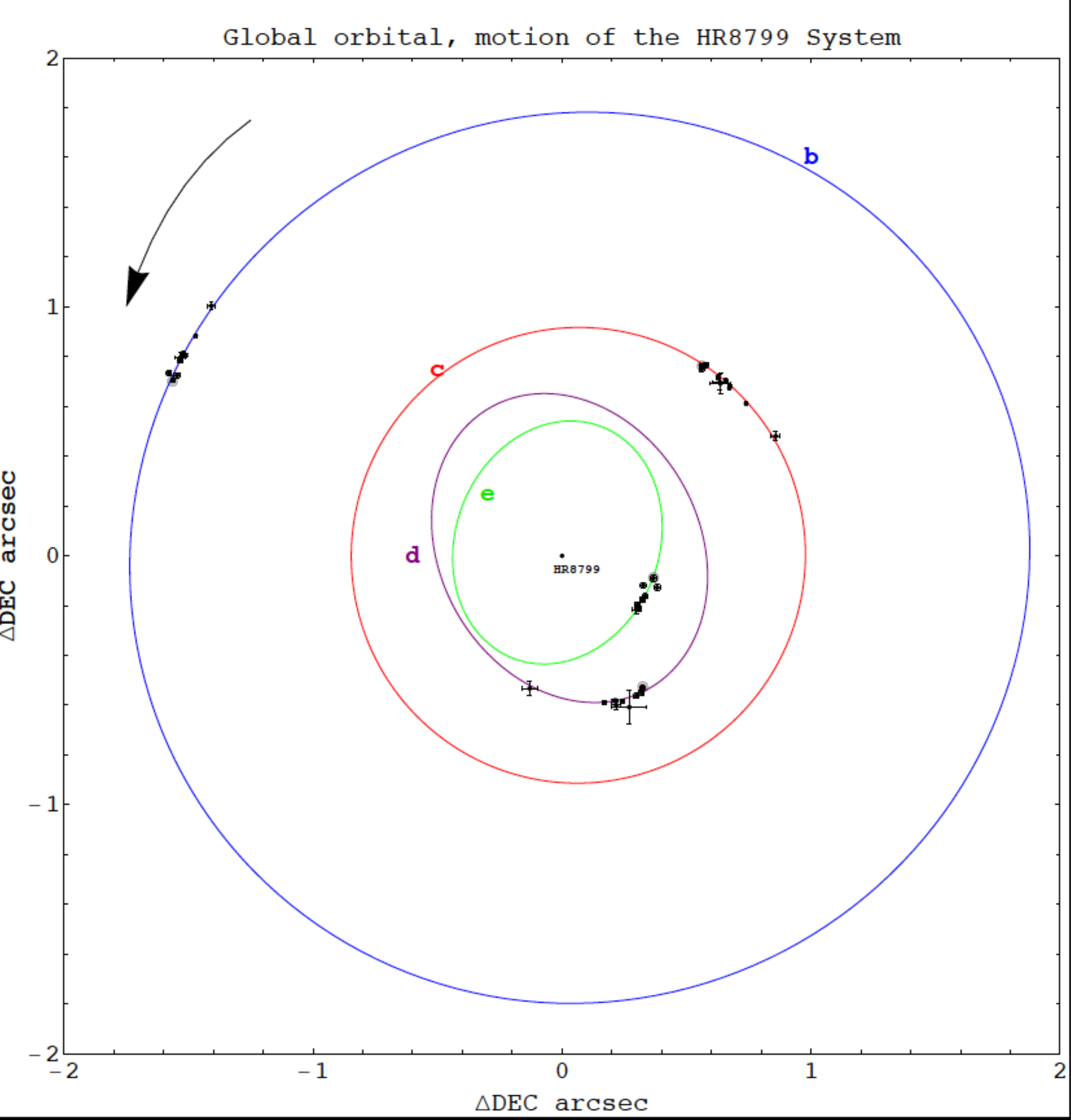}
\includegraphics[width=8 cm]{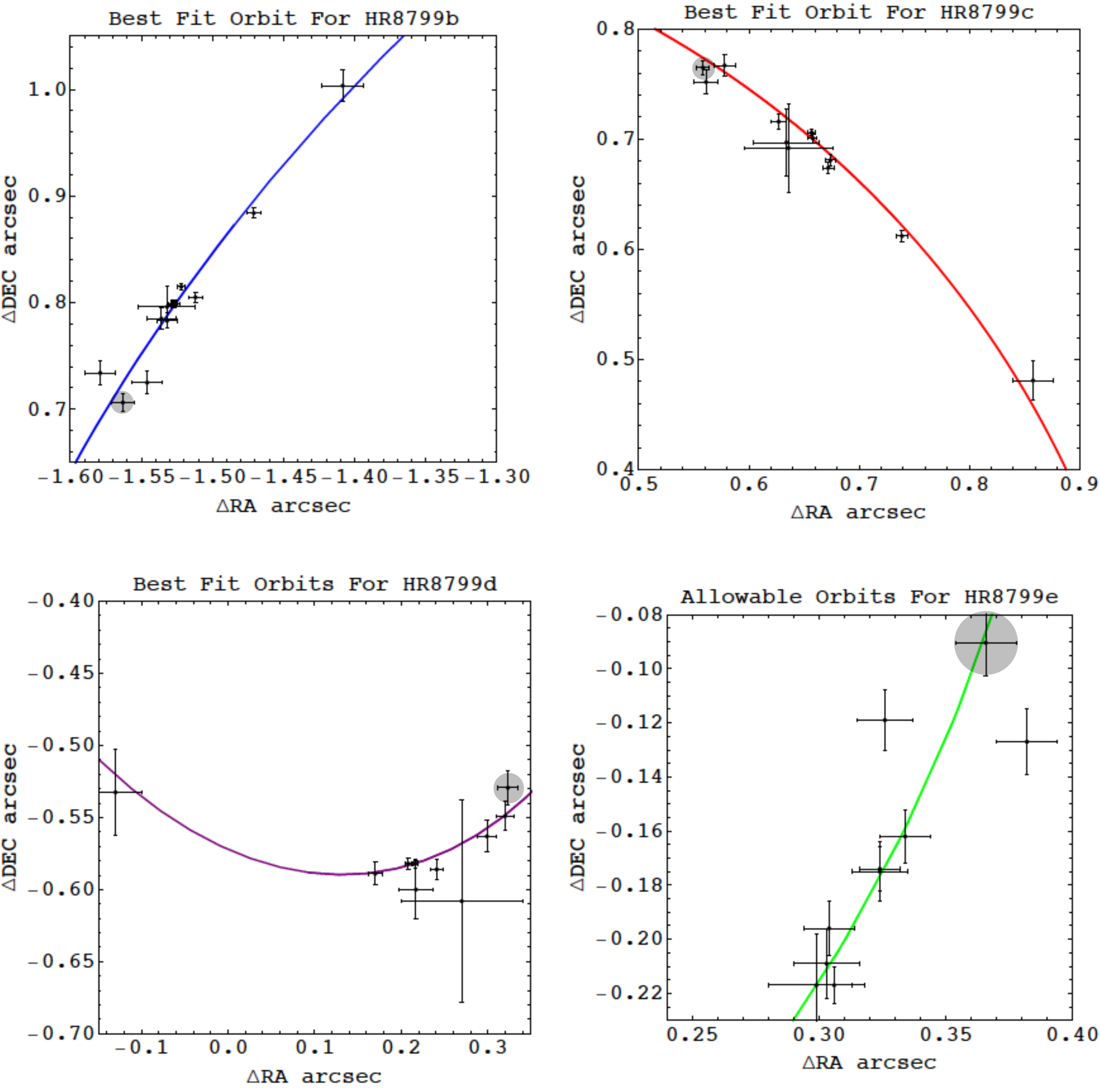}
\caption{{\bf On-sky projection of the best fit orbits for the four planets orbiting HR8799}. The right panel zooms on the portions of the orbits that have been observed in the 1998 to 2012 timespan. Note that because HR8799d's seems to be orbiting the host star in a different plane than the other three planets its on-sky trajectory appears close to the one of HR8799e than it actually is. Grey circles denote the Project 1640 epoch, which is the latest epoch considered in our analysis.}
\label{fig:HR8799FourOrbits} 
\end{figure*}
Armed with the planets' location in detector coordinates, the stellar location in the focal plane, the instrument's plate scale and position angle of absolute North we can estimate the position of the HR8799 planets relative to their host star at our epoch of observations. The uncertainties associated with the relative astrometry of our epoch are then derived as the root mean squared of the uncertainties associated with each one of these three quantities. Our astrometric estimates, and the error budget associated with them, are shown respectively on Tables \ref{tab:astromety} and \ref{tab:errorbudget}. For HR8799bc the largest uncertainty stems from our limited knowledge of stellar location in the focal plane (due to the fact that we are deriving this quantity based on natural speckles' elongation instead of using satellite spots). For HR8799de the largest uncertainty stems from the aggressiveness of the speckle suppression that necessary for a statistically significant detection. Figure \ref{fig:HR8799FourOrbits} illustrates the published relative astrometry epochs in conjunction with our P1640 epoch. The most likely orbits resulting from our orbital motion analysis, discussed next, are overlaid to these points with a different color for each planet. We will keep this color coding through the remainder of the paper when comparing the respective orbital elements of each planet one to another and constrain the orbital architecture of this system.
\begin{deluxetable}{cccc}[b!]
\tabletypesize{\scriptsize}
\tablecaption{HR8799bcde relative astrometry, June 2012.}
\tablehead{\colhead{Planet} & \colhead{Epoch} & \colhead{$\Delta RA$} & \colhead{$\delta DEC$}} 
\startdata 
HR8799b & 2012.4481 & $1.563'' ; \pm 0.005''$ & $0.706'' \; \pm 0.005''$\\
\vspace{1mm}\\
HR8799c & 2012.4481 & $-0.558'' \; \pm 0.004''$ & $0.765'' \; \pm 0.004''$\\
\vspace{1mm}\\
HR8799d & 2012.4481 & $-0.323'' \; \pm 0.006''$ & $-0.529'' \; \pm 0.006''$ \\
\vspace{1mm}\\
HR8799e & 2012.4481 & $-0.366'' \; \pm 0.006''$ & $-0.090'' \; \pm 0.006''$ \\
\vspace{1mm}
\enddata
\label{tab:astromety}
\end{deluxetable}
\begin{deluxetable}{ccccc}[h!]
\tabletypesize{\scriptsize}
\tablecaption{Astrometric error budget.}
\tablehead{\colhead{Planet} & \colhead{$\sigma_{PA}$}\tablenotemark{a} & \colhead{$\sigma_{PS}$} \tablenotemark{b}& \colhead{$\sigma_{Star}$} \tablenotemark{c}& \colhead{$\sigma_{KLIP}$} \tablenotemark{d}} 
\startdata 
HR8799b & $0.0064''$ &$0.0031''$ & $0.0065''$ & $0.0039''$\\
\vspace{1mm}\\
HR8799c & $0.0034''$ &$0.0017''$ & $0.0065''$ & $0.0039''$\\
\vspace{1mm}\\
HR8799d & $0.0022''$ &$0.0011''$ & $0.0065''$ & $0.011''$\\
\vspace{1mm}\\
HR8799e & $0.0014''$ &$0.00069''$ & $0.0065''$ & $0.011''$\\
\vspace{1mm}
\enddata
\tablenotetext{a}{$\sigma_{PA}$: uncertainty on PA offset}
\tablenotetext{b}{$\sigma_{PS}$: uncertainty on plate scale determination}
\tablenotetext{c}{$\sigma_{Star}$: uncertainty on host star location in the focal plane array}
\tablenotetext{d}{$\sigma_{KLIP}$: uncertainty stemming from residual errors induced by the KLIP reduction and the forward modeling estimator.}
\label{tab:errorbudget}
\end{deluxetable}
\begin{figure}[h!]
\center
\includegraphics[width= 1 \columnwidth]{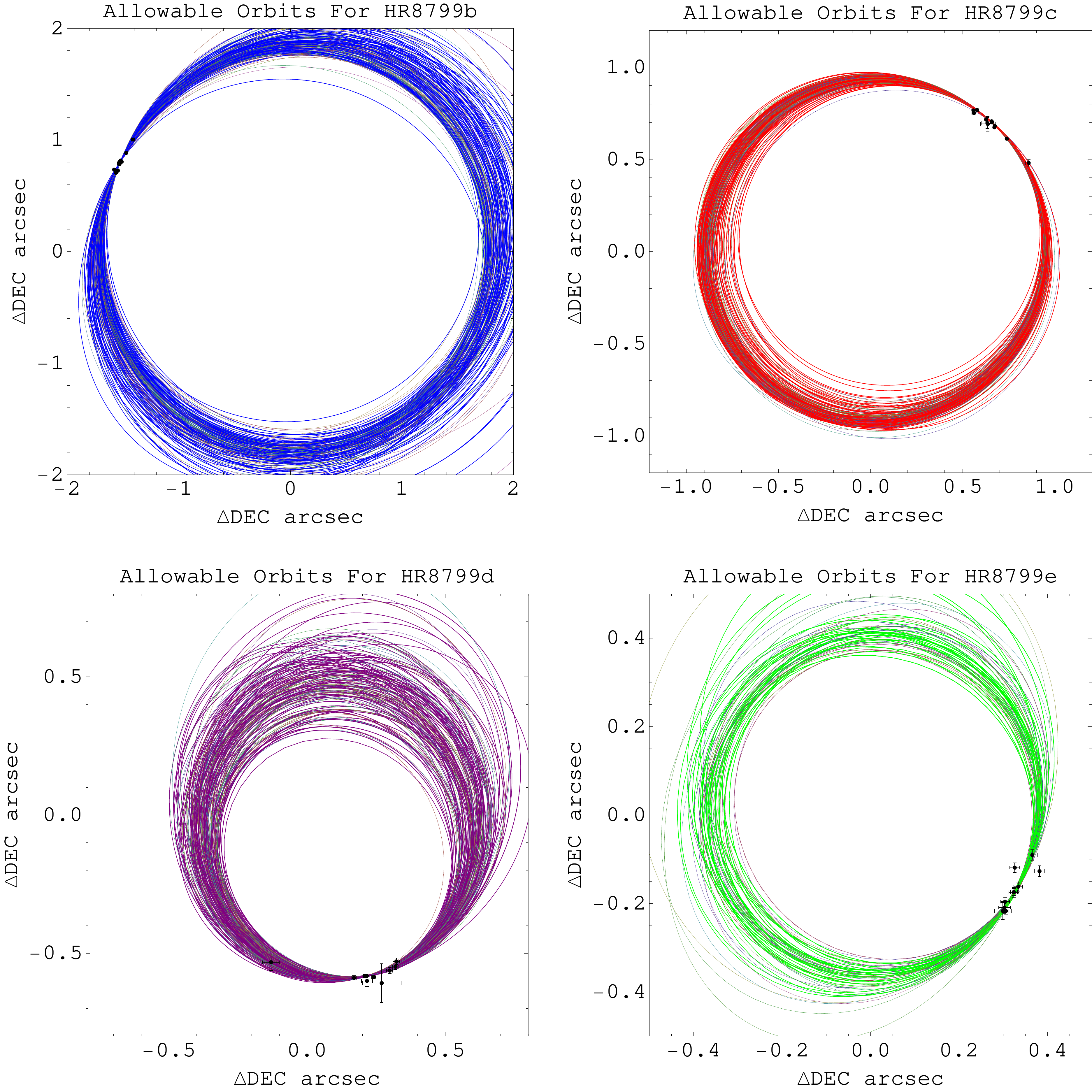}
\caption{{\bf Illustration of the on-sky projection of allowable orbits for the four planets orbiting HR8799}. Even if the orbital phase coverage is scarce due to the long orbital period of the planets our Bayesian analysis applied to the 1998 to 2012 data constrains the orbital architecture of the system.}
\label{fig:HR8799FourPlanetsAllowable} 
\end{figure}
\subsection{Orbital motion}

\begin{figure*}[t!]
\center
\includegraphics[width=19 cm]{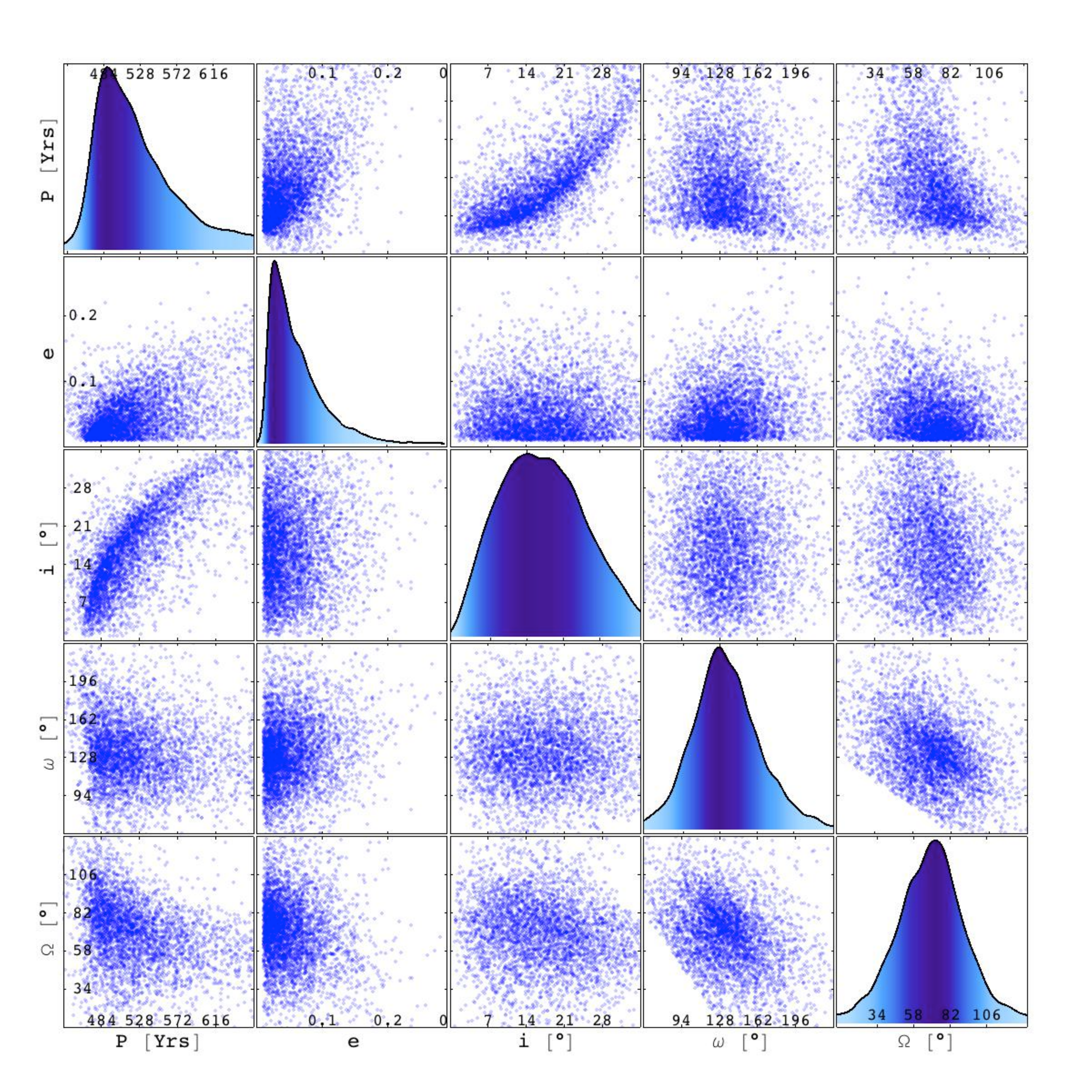}
\caption{{\bf Posterior distribution of the four Keplerian orbital elements $(P,e,i,\Omega)$ of HR8799b resulting from our Bayesian analysis of the published relative astrometry of this source with respect to its host star}. The diagonal diagrams correspond to the marginalized probability distributions and the off-diagonal ones to the correlation between various parameters.}
\label{fig:HR8799bMCMC} 
\end{figure*}

\begin{figure*}[t!]
\center
\includegraphics[width=19 cm]{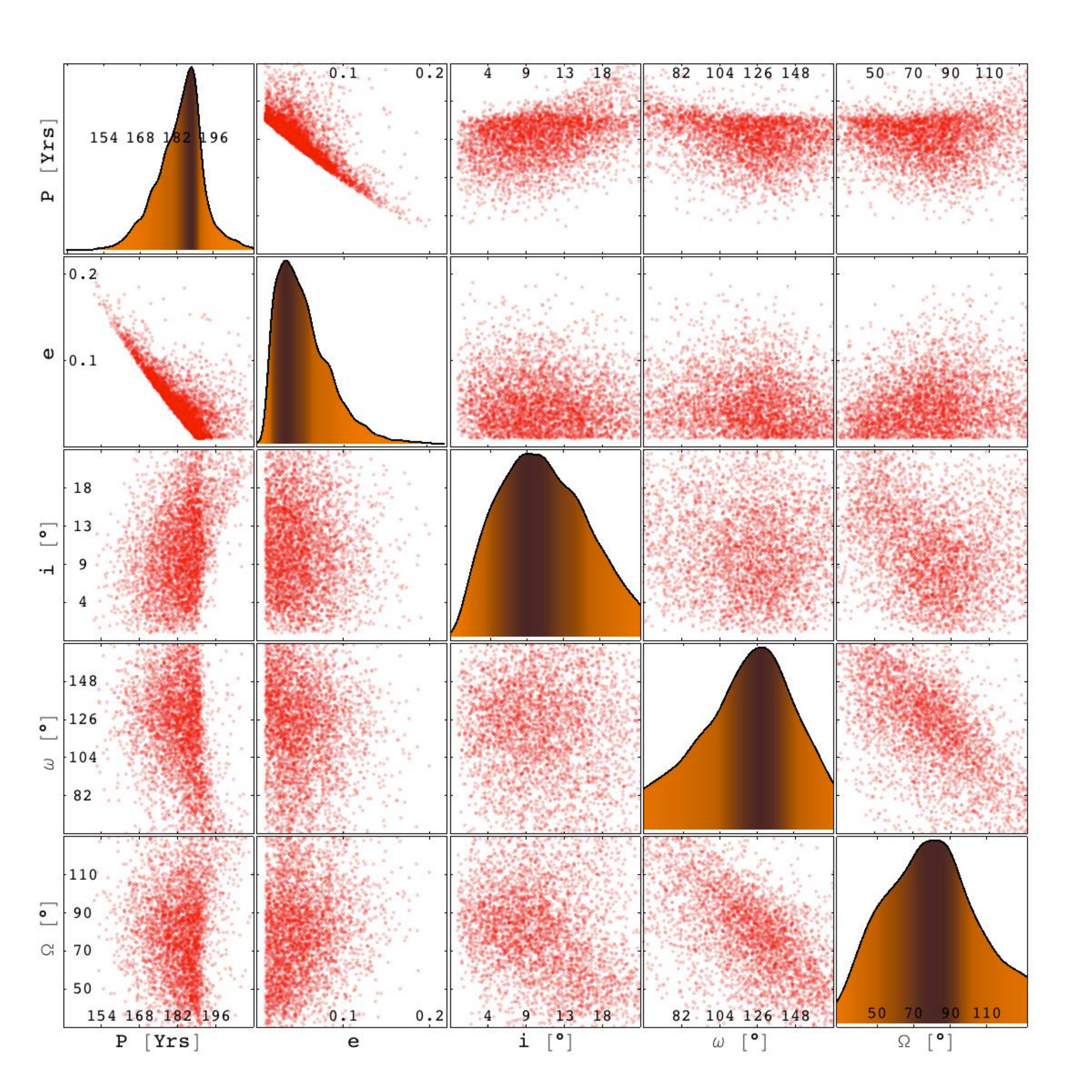}
\caption{{\bf Posterior distribution of the four Keplerian orbital elements $(P,e,i,\Omega)$ of HR8799c resulting from our Bayesian analysis of the published relative astrometry of this source with respect to its host star}. The diagonal diagrams correspond to the marginalized probability distributions and the off-diagonal ones to the correlation between various parameters.}
\label{fig:HR8799cMCMC} 
\end{figure*}

\begin{figure*}[t!]
\center
\includegraphics[width=19 cm]{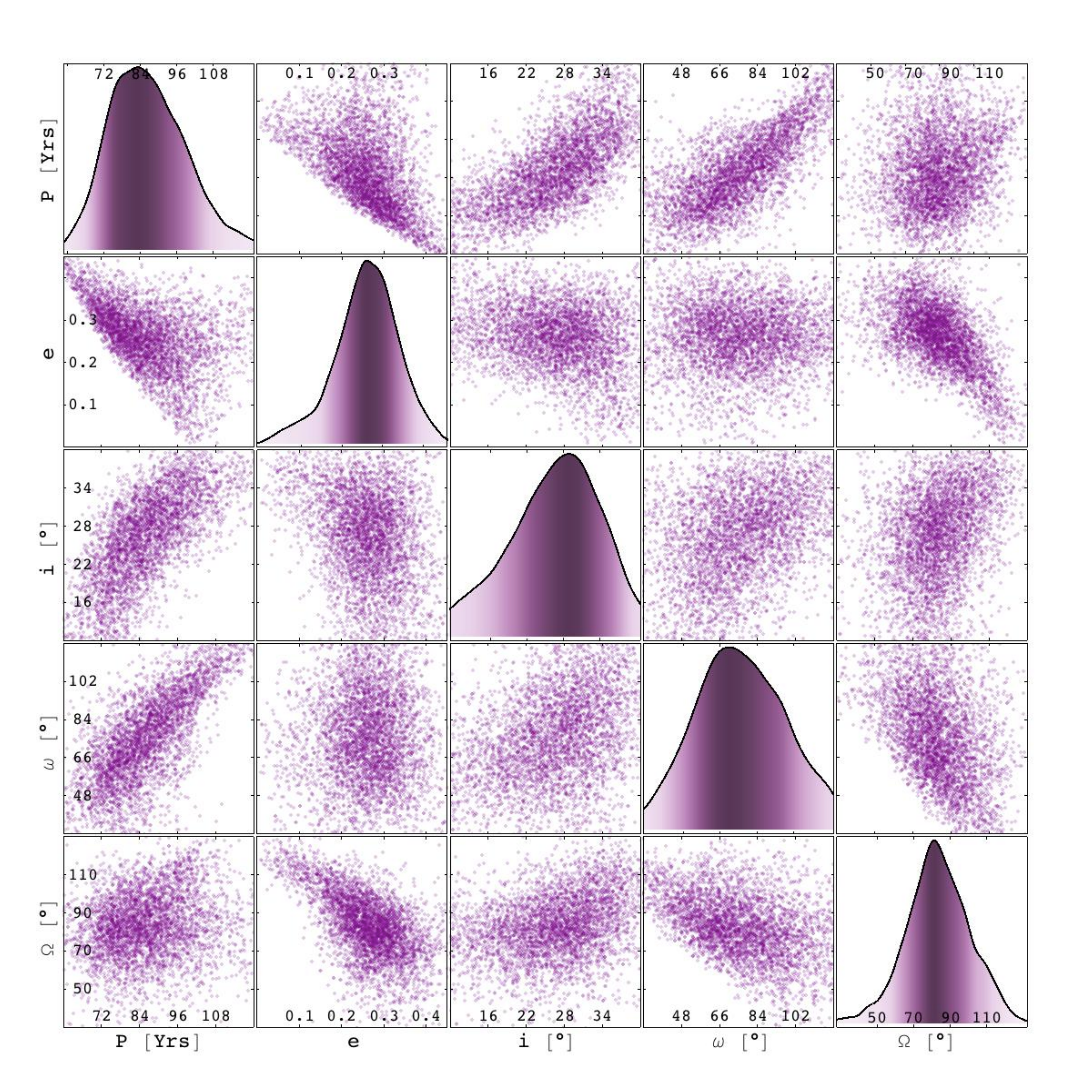}
\caption{{\bf Posterior distribution of the four Keplerian orbital elements $(P,e,i,\Omega)$ of HR8799d resulting from our Bayesian analysis of the published relative astrometry of this source with respect to its host star}. The diagonal diagrams correspond to the marginalized probability distributions and the off-diagonal ones to the correlation between various parameters.}
\label{fig:HR8799dMCMC} 
\end{figure*}

\begin{figure*}[t!]
\center
\includegraphics[width=19 cm]{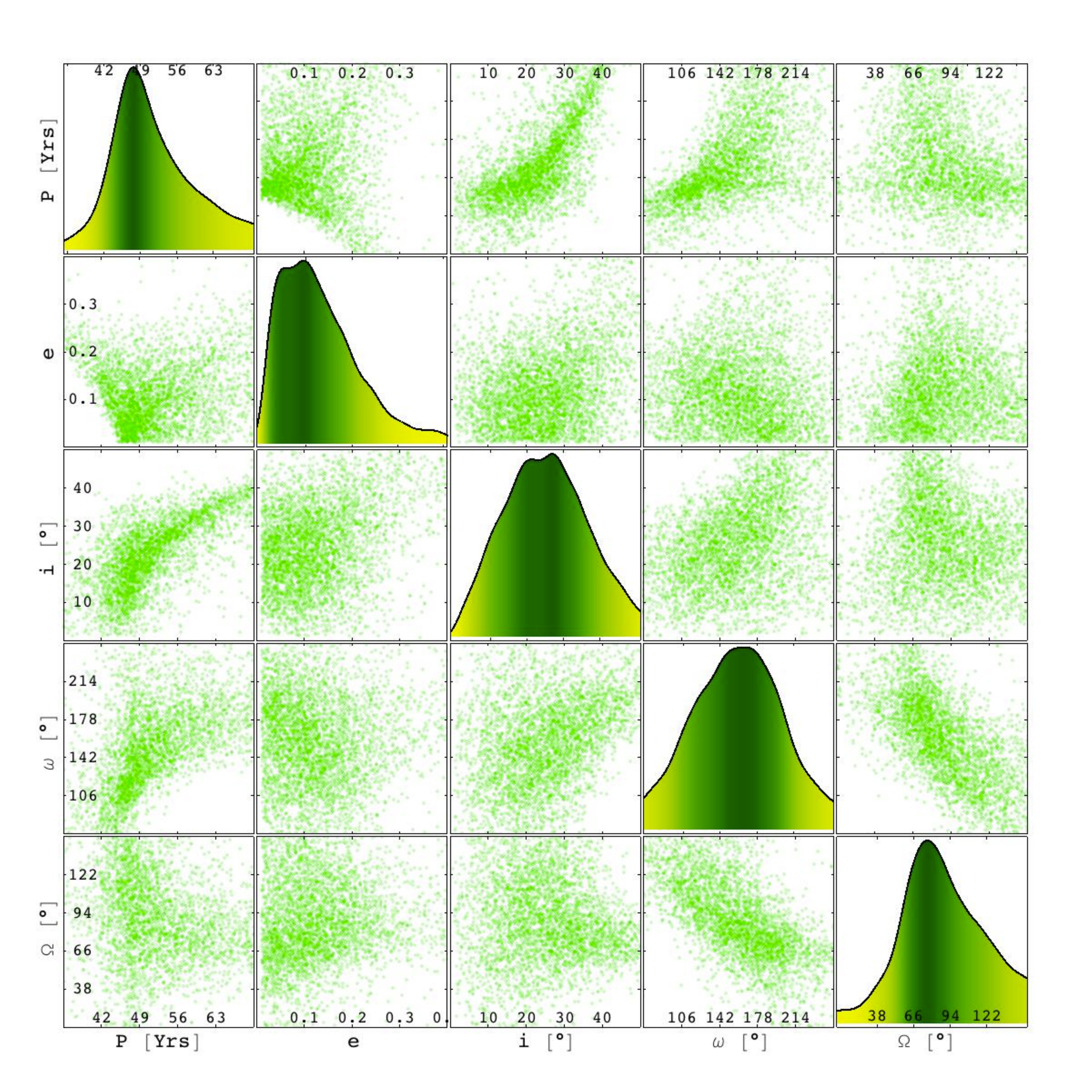}
\caption{{\bf Posterior distribution of the four Keplerian orbital elements $(P,e,i,\Omega)$ of HR8799e resulting from our Bayesian analysis of the published relative astrometry of this source with respect to its host star}. The diagonal diagrams correspond to the marginalized probability distributions and the off-diagonal ones to the correlation between various parameters.}
\label{fig:HR8799eMCMC} 
\end{figure*}

Because of HR8799 is the only directly imaged multiple planetary system to this day, the orbital motion of the four planets has been widely discussed in the literature. Published analyses of the orbital architecture of this system can be divided in two categories: non-linear least-squares fit of Keplerian elements and dynamical studies. Because of the long orbital periods and the currently limited orbital phase coverage, the parameter landscape explored by non-linear least-squares methods comprises a multitude of local minima, making it very difficult to unambiguously determine the six Keplerian elements for each planet separately. Recent papers have estimated most likely orbital architectures assuming either a set inclination for the four planets \citep{lmd09,2011A&A...528A.134B,2013A&A...549A..52E}, or coplanar and locked in mean motion resonances orbits \footnote{Note that \citet{2012ApJ...755L..34C} also conducted an orbital motion analysis without any resonant assumption, and naturally obtained looser constraints on Keplerian elements than when assuming a Laplace mean motion resonance} \citep{Sou11,2012ApJ...755L..34C}. On the other hand dynamical analysis can constrain the dynamical mass of the planets (upon finding orbital architectures stable over durations at least as long as the estimated stellar age) and can predict near future orbital position \citep{fm10,rks09,gm09,2013A&A...549A..52E,2013arXiv1308.6462G,mzk10}. However, dynamical models are generally used in conjunction with strong assumptions regarding coplanarity and mean motion resonances. It was recently shown, under such assumptions, that one could also include planetary migration mechanisms into a dynamical analysis and thus deliver joint information regarding the planets' formation history and masses \citep{2013arXiv1308.6462G}. When invoking mean motion resonances to stabilize the system most authors identified the Laplace 1:2:4:8 resonance as a promising architecture that is compatible with both the available astrometric epochs and masses in the planetary regime ($5-10 \; M_{Jup}$). The objective of the present paper is to complement both approaches, either non-linear least squares fit or dynamical studies, by answering the following question {\em ``what is the most likely set of Keplerian elements for each planet in the HR8799 system given the data at hand from 1998 to 2012?''}. To do so we carry out to a Bayesian analysis of the published astrometric epochs. Since our approach does not need to resort to any assumptions about the architecture of the system (coplanarity in particular) or the planetary masses it should provide a good empirical baseline to test published dynamically favorable architectures. We also seek to complement more ``data oriented'' methods that either need strong assumptions on the orbital architecture of the system or do not take full advantage of the tools provided by Bayesian inference \citep{Sou11,2012ApJ...755L..34C,lmd09,2011A&A...528A.134B,2013A&A...549A..52E}.

Bayesian inference using Markov Chain Monte Carlo methods has been extensively used for the detection and characterization of exo-planets using indirect methods: radial velocity \citep{2005AJ....129.1706F,2006ApJ...642..505F,2011MNRAS.410...94G,2012ApJ...745..198H}, transits \citep{2013PASP..125...83E} and gravitational micro-lensing \citep{2011ApJ...738...87S}. Recently the direct imaging community has focused on similar methods to characterize the orbits of Beta Pictoris b \citep{2012A&A...542A..41C} and Fomalhault b \citep{2013ApJ...775...56K}. We conducted our analysis of the orbital architecture of HR8799 using two difference MCMC samplers: the Metropolis Hastings algorithm discussed in \citet{2006ApJ...642..505F,2012A&A...542A..41C,2013ApJ...775...56K} and the Affine Invariant Sampler described in \citet{2013PASP..125..306F}. Both approaches did yield almost identical posterior distributions for the six orbital elements of each planet. We direct the reader to the aforementioned publications for detailed description of our methodology and below we only describe the broad lines of our Bayesian analysis:
\begin{itemize}
\item[-] We use all the published epochs summarized in \citet{2013A&A...549A..52E} augmented by our P1640 points. 
\item[-] For each planet we seek to constrain the six orbital elements period $P$, eccentricity $e$, inclination $i$, longitude of ascending node $\Omega$, argument of periastron $\omega$, epoch at periastron $t_{P}$. We consider the following state vector $\mathbf{x} = (log(P),e,\cos i,\omega+\Omega,\omega-\Omega,t_{p}$). 
\item[-] Case of the Metropolis Hasting Sampler: Our prior distributions of periods for each planet are uniform, centered around the``circular face on period'' ($P_{cfo}$ not to be confused with the commonly eccentric orbit $P_{ceo}$) derived from the discovery epochs and covers more than a full decade around that value $P_{prior} \in [P_{cfo}/4 , 4 P_{cfo}]$. Since high eccentricities have been ruled out \citep{mzk10} we initialize our chains according to a uniform distribution spanning $[0,0.8]$. We assume a uniform prior distributions for $\cos i \; \in [-1,1]$, $\omega \; \in [0,2 \pi]$, $\Omega \; \in [0,2 \pi]$ and $t_p \in [P_{cfo}- 4 P_{cfo},P_{cfo}+4 P_{cfo}]$. We improve convergence by using the transformation $\mathbf{u}(\mathbf{x})$ described in Appendix A of \citet{2012A&A...542A..41C} and use for $t_0$, the origin of mean anomalies \citep{2006ApJ...642..505F}, the epoch with the largest number of contemporaneous observations (usually the epoch at which confirmation of physical association was unambiguously established). Since the orbital phase coverage is much smaller than in the case of Beta Pictoris b and more similar to Fomalhault b we further improve convergence by using a parallel tempering ladder \citep{2005blda.book.....G,2013ApJ...775...56K}. 
\item[-] Affine Invariant Sampler: we follow the recommendation of \citet{2013PASP..125..306F} and we first initialized our walkers in a small ball centered on the most likely orbital elements derived using the Metropolis Hastings sampler. We then take advantage of the enhanced computational speed of the Affine Invariant Sampler and explore a variety of walker initialization points among the range discussed above. We verify that the better behaved chains in terms of acceptance rate, autocorrelation and overall chi-squared are indeed the ones corresponding to the most likely orbital elements estimated using the Metropolis Hasting sampler. We use these latter chains for inference.
\item[-] Both the small orbital phase coverage and the high star/planet mass ratios do not allow us to take advantage of our inference chains to carry out the class dynamical mass estimates described in \citet{2009ApJ...706..328D,2010ApJ...711.1087K,cjf12}. Since we cannot reach this level of precision we do not marginalize over the distance to HR8799 (assumed to be $36.4$ pc) nor the host stellar mass (assumed to be $1.51 \; M_{Sun}$).
\end{itemize}

Our results for each planet are shown on Figures \ref{fig:HR8799bMCMC} to \ref{fig:HR8799eMCMC}. The elements of the ``most likely orbit'' and 1-$\sigma$ confidence intervals indicated are summarized in Table \ref{tab:keplerian} and the orbits associated with the lowest $\chi^2$ value are displayed on Figure \ref{fig:HR8799FourOrbits}. Finally the ensemble of allowable orbits, located in the 1-$\sigma$ confidence intervals is shown in Figure \ref{fig:HR8799FourPlanetsAllowable}, this illustrates the degeneracies associated with fitting Keplerian orbital elements using astrometric data spanning only a small portion of orbital phase. Nevertheless, even in the presence of such degeneracies, the marginalized probability density function in Figure \ref{fig:HR8799bMCMC} to Figure \ref{fig:HR8799eMCMC} can help us to significantly constrain the architecture of the HR8799 system. We discuss these aspects in Section \ref{sec:Discuss}. In a future paper we will test the dynamical stability of this ensemble of allowable orbits or order to obtain a more precise understanding of the orbital architectures of system and of the planet's dynamical masses.

However before delving into our interpretation of Figures \ref{fig:HR8799bMCMC} to \ref{fig:HR8799eMCMC} we remind the reader of the caveats associated with our Bayesian analysis of the astrometric history of this system. Markov Chains Monte Carlo are known to be very sensitive to underestimated systematic biases (e.g not captured by published error bars) in the various astrometric measurements. The astrometric data supporting the analysis of \citet{2012A&A...542A..41C} and \citet{2013ApJ...775...56K} was extremely homogenous as it was based on two instruments at most, and all the imaging data reduction had been conducted by a single team. On the contrary the astrometric history underlying our analysis is heterogeneous and comprises estimates stemming from at least six observatories and various data analysis approaches. For instance, should small discrepancies between absolute North calibrations, beyond the reported error bars, occur between observatories then the posterior distributions in Figure \ref{fig:HR8799bMCMC} to Figure \ref{fig:HR8799eMCMC} will be biased. This could for instance be the source of the minor differences between our results and the confidence intervals reported in \citet{2012ApJ...755L..34C}. While our results for HR8799bcd are generally in good agreement with \citet{2012ApJ...755L..34C} some of our confidence intervals do not overlap. The presence of biased observations in the astrometric history of HR8799 combined with the sensitivity of MCMC approaches to such unaccounted uncertainties could be the sources of these discrepancies. Alternatively it could be that the approach in \citet{2012ApJ...755L..34C} does not sufficiently explore the $\chi^2$ landscape. Homogenous observations with new generation Extreme Adaptive Optics instruments, which posses superior astrometric accuracy, will largely resolve this issue over the upcoming decade. Note however that systematics between observatories (and epochs if necessary) could be included in the MCMC state vector, just as mean radial velocities are included in observations of the reflex motion of exo-planetary host stars along the line of sight \citep{2006ApJ...642..505F}. Should this approach be carried out, it would then yield confidence intervals for the architecture of directly imaged exo-planetary system that are free of systematics  (and thus more reliable). While this would be an extremely interesting academic exercise, including potential systematics in our analysis is beyond the scope of the present paper.
%
% \begin{deluxetable}{ccccc}[b!]
% \tabletypesize{\scriptsize}
% \tablecaption{Most likely Keplerian elements for the planets in the HR8799 system and confidence interval.}
%\tablehead{\colhead{} & \colhead{HR8799b} & \colhead{HR8799c} & \colhead{HR8799d} & \colhead{HR8799e}} 
%\startdata 
%\vspace{0.5mm}\\
% $P_{\chi^2_{min}}$ [Yrs] & 558.2 & 237 & 85.8 & 64.0\\
%$P, \; 1\sigma$ & [501.6,613.9]& [201.1,276.8] & [73.7,97.0] & [52.0,77.0]\\
%\vspace{0.5mm}\\
%\hline
%\vspace{0.5mm}\\
% $<P>$ [Yrs] & 538.2 & 237 & 85.8 & 64.0\\
%$P, \; 1\sigma$ & [501.6,613.9]& [201.1,276.8] & [73.7,97.0] & [52.0,77.0]\\
%\vspace{0.5mm}\\
%\hline
%\vspace{0.5mm}\\
% $e_{\chi^2_{min}}$ & 0.079 & 0.18 & 0.11 & 0.16 \\
%$e, \; 1\sigma$ & [0.012,0.16] & [0.10,0.27] & [0.028,0.22] & [0.03,0.28]\\
%\vspace{0.5mm}\\
%\hline
%\vspace{0.5mm}\\
% $i_{\chi^2_{min}} \; [^{\circ}]$ & 19.7& 19.8 & 25.56 & 25.1 \\
%$i, \; 1\sigma$ & [12.9,26.33] & [12.8,26.4] & [19.0,32.1] & [15.1,35.4]\\
%\vspace{0.5mm}\\
%\hline
%\vspace{0.5mm}\\
% $\Omega_{\chi^2_{min}} \; [^{\circ}]$ & 88.4 & 85.5 & 97.1 & 84.4\\
%$\Omega, \; 1\sigma$ & [84.2,92.5] & [83.1,87.9] & [93.9,99.8] & [78.5,90.9] \\
%\vspace{0.5mm}\\
%\enddata
%\label{tab:keplerian}
%\end{deluxetable}

\begin{deluxetable}{ccccc}[b!]
\tabletypesize{\scriptsize}
\tablecaption{Most likely Keplerian elements for the planets in the HR8799 system and confidence interval.}
\tablehead{\colhead{} & \colhead{HR8799b} & \colhead{HR8799c} & \colhead{HR8799d} & \colhead{HR8799e}} 
\startdata 
\vspace{0.5mm}\\
$P_{\chi^2_{min}}$ [Yrs] & 525.3 & 174.5 & 87.4 & 58.9\\
$P, \; 1\sigma$ & [479.5,574.9]& [164.1,184.9] & [74.5,99.8] & [45.1,70.3]\\
\vspace{0.5mm}\\
%\hline
%\vspace{0.5mm}\\
%$a_{\chi^2_{min}}$ [AU] & 77.8 & 43.9 & 22.2 & 18.4\\
%$a, \; 1\sigma$ & [72.4,82.9]& [39.4,48.7] & [20.2,24.2] & [16.0,20.8]\\
%\vspace{0.5mm}\\
\hline
\vspace{0.5mm}\\
$e_{\chi^2_{min}}$ & 0.056 & 0.086 & 0.26 & 0.14 \\
$e, \; 1\sigma$ & [0.018,0.092] & [0.042,0.12] & [0.18,0.33] & [0.045,0.21]\\
\vspace{0.5mm}\\
\hline
\vspace{0.5mm}\\
$i_{\chi^2_{min}} \; [^{\circ}]$ & 17.2& 10.5 & 26.3 & 25.5 \\
$i, \; 1\sigma$ & [8.5,25.9] & [5.1,16.0] & [18.2,34.2] & [13.3,37.5]\\
\vspace{0.5mm}\\
\hline
\vspace{0.5mm}\\
$\omega_{\chi^2_{min}} \; [^{\circ}]$ & 134.1 & 123.9 & 76.4 & 160.7\\
$\omega, \; 1\sigma$ & [102.8,164.9] & [100.1,147.5] & [53.1,99.9] & [111.7,206.0] \\
\vspace{0.5mm}\\
\hline
\vspace{0.5mm}\\
$\Omega_{\chi^2_{min}} \; [^{\circ}]$ & 69.4 & 81.4 & 82.9 & 89.0\\
$\Omega, \; 1\sigma$ & [46.5,91.1] & [60.2,101.1] & [66.8,100.4] & [58.3,128.3] \\
\vspace{0.5mm}\\
\enddata
\label{tab:keplerian}
\end{deluxetable}

\section{Discussion}
\label{sec:Discuss}

\subsection{Coplanarity}
\begin{figure}[t!]
\center
\includegraphics[width= 0.69 \columnwidth]{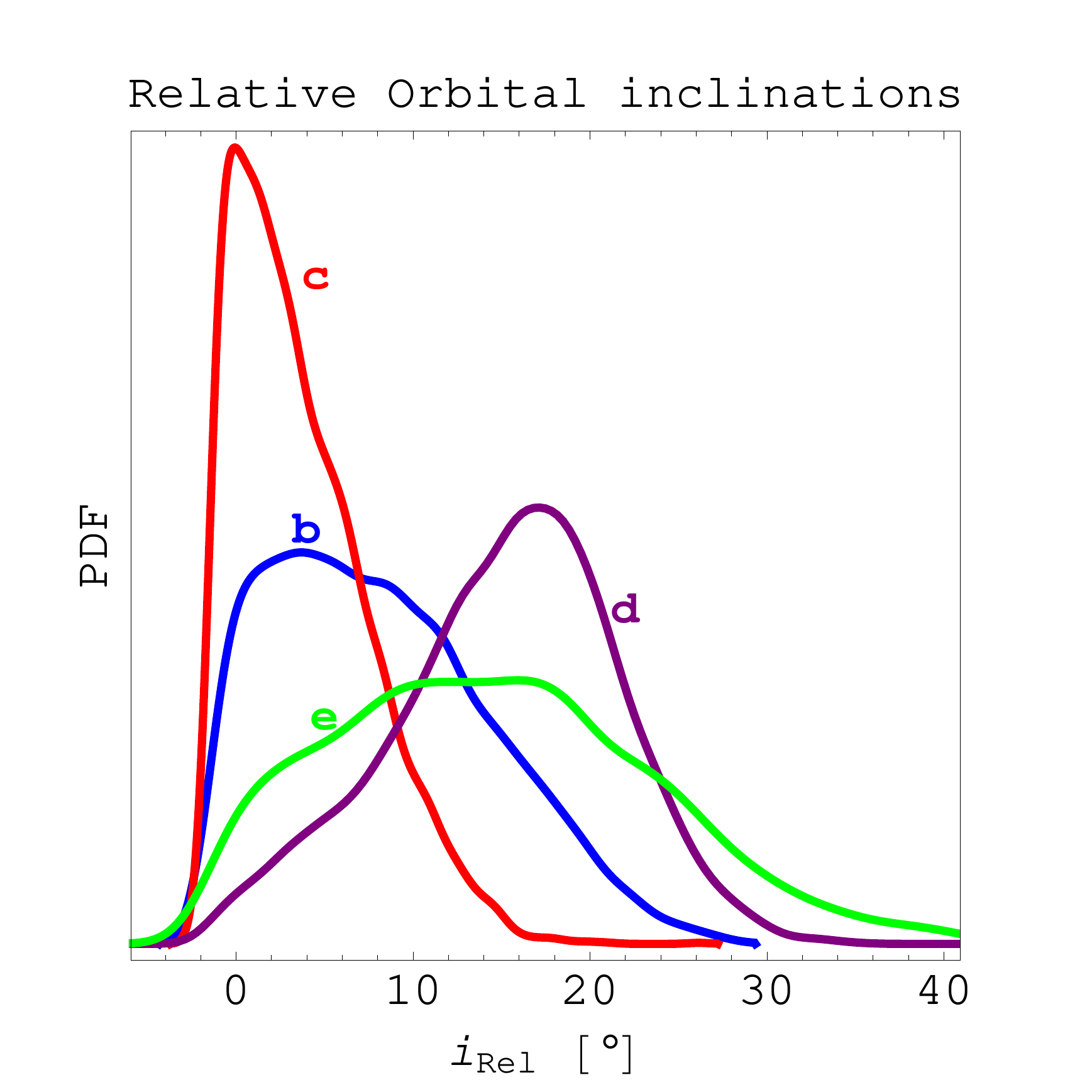}
\caption{{\bf coplanarity test for the four planets in the HR8799 system}: this figure show the relative inclinations of the orbital planes with respect to the most likely (whereas the inclinations in Figures \ref{fig:HR8799bMCMC} to \ref{fig:HR8799eMCMC} are with respect to the line of sight). While this figure does not unambiguously rule out coplanarity, it suggests that HR8799bc most likely orbit in the same plane while HR8799d's orbit is out of the plane. More data is needed to constrain the plane of the orbit of HR8799e.}
\label{fig:AnglesHistogram} 
\end{figure}

In order to test coplanarity we folded the inclinations and longitudes of ascending notes on Figure \ref{fig:HR8799bMCMC} to \ref{fig:HR8799eMCMC} to compute the direction of the vector orthogonal to the orbital plane of each realization of the orbits of HR879bcde. We then used as a reference the direction of the vector orthogonal to the orbital plane of the most likely orbit of HR8799c. Figure \ref{fig:AnglesHistogram} shows the relative inclinations of each planet with respect to this reference. While this figure does not unambiguously rule out coplanarity, it suggests that HR8799bc most likely orbit in the same plane while HR8799d's orbit is out of the plane. More data is needed to constrain the plane of the orbit of HR8799e. This non coplanar orbital architecture has not been included in recently published dynamical analyses. As a final sanity check we further test the robustness of our analysis regarding the orbits of HR8799d. A direct inspection of Figure \ref{fig:HR8799FourOrbits} hints that the out of plane best fit for HR8799d might stem from a biased astrometry in the 1998 HST-NICMOS epoch \citep{Sou11} since the large temporal lever arm it is responsible for strong constraints on the orbit. We tested the robustness of our results to this bias by removing the HST-NICMOS point from the astrometric history. This tests did yield posterior distributions of orbital elements similar to Fig \ref{fig:HR8799dMCMC} (albeit with larger uncertainties in the period when removing the 1998 epoch). We can thus rule out this scenario and conclude that, in the absence of other unidentified pathological cases, the orbit of HR8799d is misaligned by $\sim 15 - 20^{\circ}$ compared to the roughly coplanar HR8799bce orbits. 

%As discussed below this result translates into eccentricity distributions and period ratios which do not match the dynamically stable orbital architecture published so far. We will revisit this problem in light of our new analysis in a subsequent paper \cite{AArronDynamocal}.
%
\subsection{Eccentricities}
\begin{figure}[h!]
\center
\includegraphics[width= 0.7 \columnwidth]{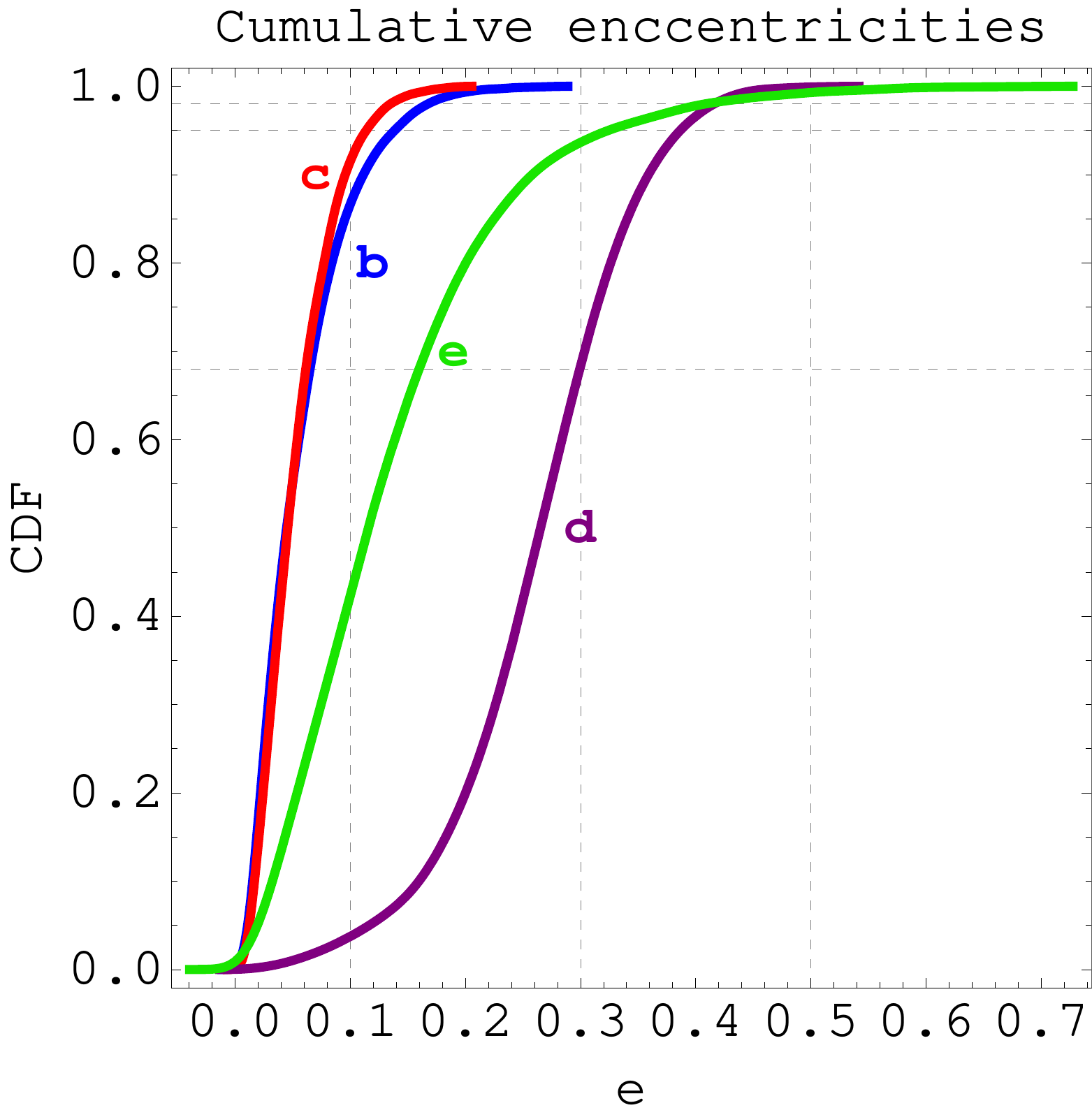}
\caption{{\bf Cumulative distributions of the eccentricities of the four planet orbiting HR8799}: the dashed horizontal lines, from bottom to top, correspond to the $68 \%$, $95 \%$ and $98 \%$ confidence levels. Our analysis of the current astrometric history confirms the published eccentricity hierarchy in other works with HR8799d being the most eccentric planet and HR8799bc featuring almost circular orbits ($e<0.1$ with $68 \%$ confidence). More orbital coverage will be necessary to firmly establish the eccentricity of HR8799e.}
\label{fig:EccentricitiesCumulative} 
\end{figure}

All four planets appear to orbit HR8799 with low eccentricities, which has been predicted using dynamical arguments. Indeed a circularizing mechanism ought to have occurred in the youth of this system in order for it to have lasted a few tens of million years \citep{fm10}. Figure \ref{fig:EccentricitiesCumulative} shows the cumulative distribution of eccentricities in the HR8799 system, with horizontal lines denoting the $68 \%$, $95 \%$ and $98 \%$ confidence levels. The overall eccentricity distribution is consistent with other studies e.g. by \citet{Sou11,rks09,2013arXiv1308.6462G}, which identified HR8799d as the most eccentric planet: the 1-$\sigma$ upper limit for $e_b,e_c,e_e$ are respectively $0.07,0.06,0.12$ while the same upper limit of HR8799d is $0.3$. Thus in spite of the non-coplanarity discussed above, our analysis of the astrometric history of this system hints that HR8799d seems to have a special role in the eccentric hierarchy. The fact that HR8799d orbits at a larger eccentricity and a likely off-plane inclination might be the fingerprint of dynamical interactions during the formation of this planetary system. 
\subsection{Mean Motion resonnances}

\begin{figure*}[htb]
\center
\includegraphics[height = !,width=17cm]{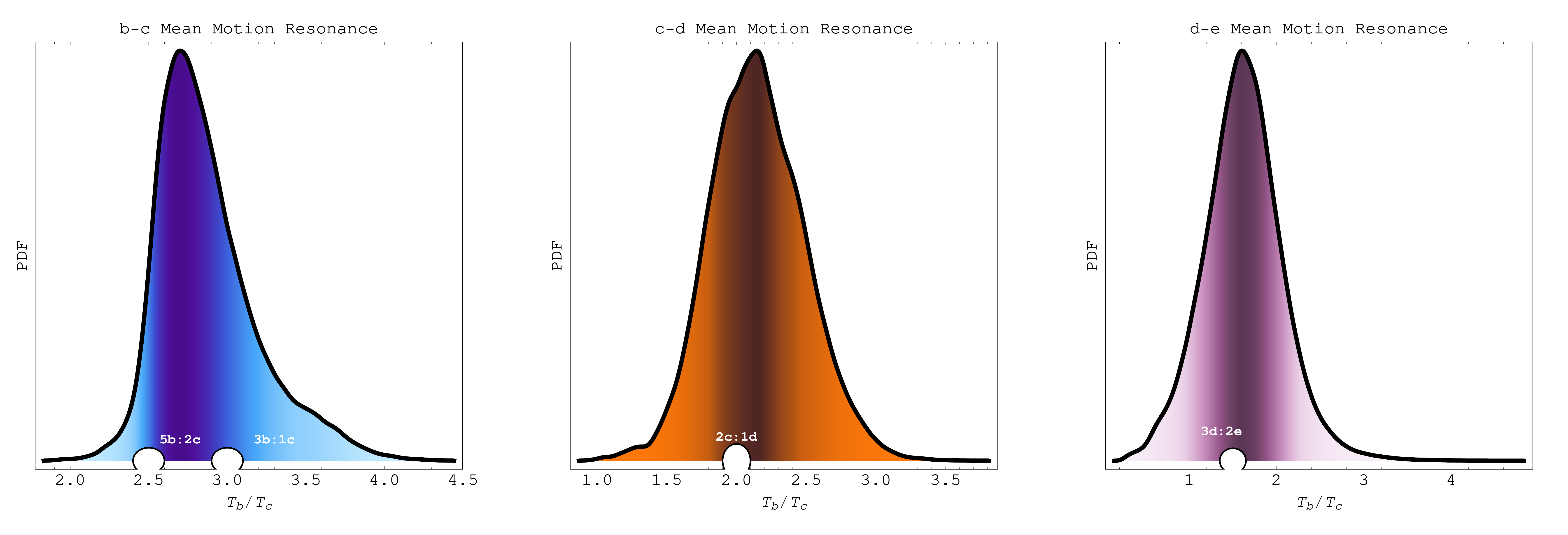}
\caption{{\bf Left: mean motion resonance test for HR8799b and c}. Histogram of the period ratio $T_b/T_c$. The lowest order mean motion resonances which appear compatible with our analysis of the current astrometric history of the HR8799 system is $3b:1c$ and the $2b:1c$ is ruled out at a high level. {\bf Bottom: mean motion resonance test for HR8799c and d}: histogram of the period ratio $T_c/T_d$. The lowest order mean motion resonances which appear compatible with our analysis of the current astrometric history of the HR8799 system is the $2:1$ resonance previously identified in the literature. {\bf Middle: mean motion resonance test for HR8799c and d}: histogram of the period ratio $T_c/T_d$. The lowest order mean motion resonances which appear compatible with our analysis of the current astrometric history of the HR8799 system is the $2:1$ resonance previously identified in the literature. {\bf Right: Mean motion resonance test for HR8799d and e}. Histogram of the period ratio $T_d/T_e$. Our analysis of the current astrometric history of the HR8799 system appears to favor a $3d:2e$ resonance but does allow rule out the $2d:1e$ which has been previously suggested in the literature. } 
\label{fig:MMRs} 
\end{figure*}

Mean motion resonances play a crucial role in stabilizing the HR8799 system as first pointed out by \citet{fm10}. While resonances are not strictly limited to integer period ratios, we identify on Figures \ref{fig:MMRs} the lowest order ratios that are compatible with our analysis of the astrometric history of the system:
\begin{itemize}
\item {\bf b-c resonance}: our analysis seems to favor a period ratio between 5b:2c and 3b:1c. Interestingly the 2b:1c resonance which had been identified as a promising candidate by dynamical studies assuming coplanarity between the two objects can be ruled out by our analysis. 
\item {\bf c-d resonance}: our analysis seems to favor period ratios commensurate with a 2c:1d resonance. 
\item {\bf d-e resonance}: our analysis yields a period ratio histogram centered around a 3:2e resonance but is also compatible with a 2d:1e. Further astrometric monitoring of HR8799e is required to disentangle these scenarios. 
\end{itemize}
Thus even if HR8799d most likely does not orbit in the same plane as HR8799c and e, the period ratios involving these planets do not rule out a 1e:2c:4d Laplace resonance. They however favor a 3d:2e resonance for the inner pair of planets. The period ratio for the two outer planets favored by our analysis does not suggest a Laplace resonance but is consistent with \citet{2012ApJ...755L..34C}. Since resonances are a strong mechanism to stabilize multiple planetary systems the confidence intervals derived in the present paper can serve as a good first guess to study secular dynamical interaction in the HR8799 system. Future N-body analyses of this system should thus be carried out to complement our work and further constrain the period ratios discussed above. 

\subsection{Likelihood of close encounter.}

The tight bounds on the dynamical mass of HR8799bcde can potentially be estimated via N-body simulations over the lifetime of the system for the ensembles of allowable Keplerian elements derived by our analysis. However this exercise is beyond the scope of the present paper and we will devote a future communication to this matter \citep{AArronDynamocal}. Here we nevertheless illustrate how such an exercise could be used to bound the dynamical mass is planets in this system based on our results in Figures \ref{fig:HR8799bMCMC} to \ref{fig:HR8799eMCMC}. Our method consists of calculating the fraction of allowable orbits that pass the ``close encounter test": following the arguments of \citet{cfm08} we define as a close encounter an epoch for which the two planets are within four mutual Hill radii. For each object we draw a set of 1000 Keplerian elements according to the posterior distributions of our Bayesian analysis. We then calculate the position of each planet in a three dimensional frame centered on the star over the next $10^{3}$ years ($\sim$ twice the period of HR8799b). For each neighboring objects (b-c, c-d and d-e) we compute the distance between the $\sim 10^6$ pairs of outer-inner orbits, vary the mass of the objects and estimate the percentage of orbit's combinations that do not feature a close encounter as defined above. 

\begin{figure*}[t!]
\center
\includegraphics[height = !,width=20cm]{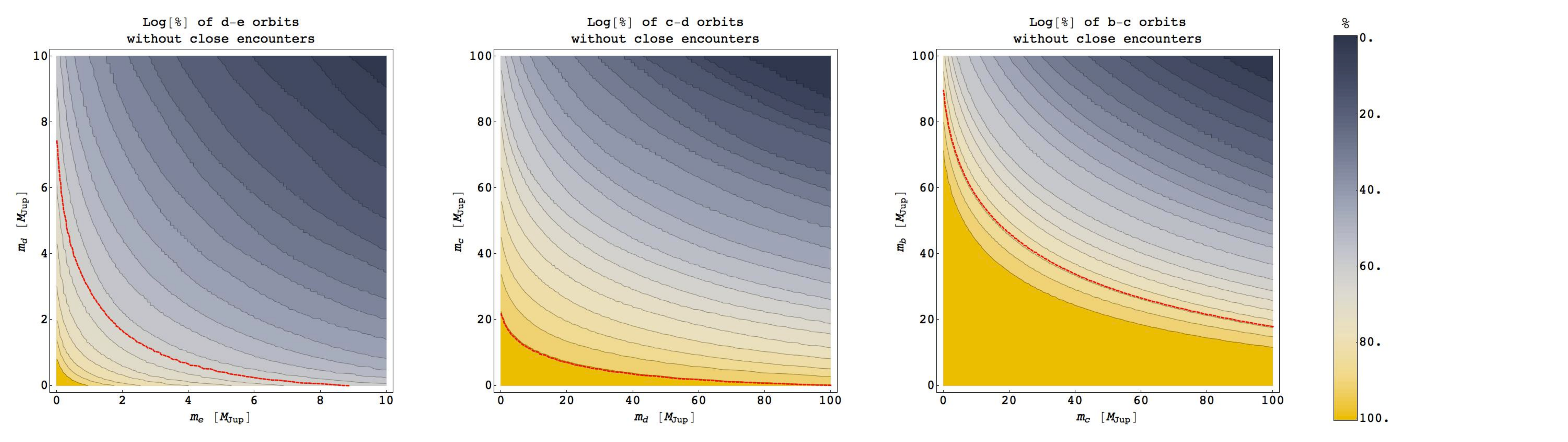}
\caption{{\bf Left: zeroth order dynamical stability test of HR8799d and e}. Percentage of the orbits allowed by our analysis of the current astrometric history of the HR8799 system which preclude a ``close encounter'' with the next orbital period of HR8799d as a function of each planet's masses. We indicate in red the boundary of the domain where more than $68 \%$ of the combinations of our allowable orbits does not feature a ``close encounter". While this is rough test from a dynamical standpoint the absence of any orbit pairs without close encounters for masses larger than $12 \; M_{Jup}$ indicates that most likely HR8799d and e dynamical masses lie below this threshold. {\bf Center: zeroth order dynamical stability test of HR8799c and d}. Percentage of the orbits allowed by our analysis of the current astrometric history of the HR8799 system which preclude a ``close encounter'' with the next orbital period of HR8799c as a function of each planet's masses. {\bf Right: zeroth order dynamical stability test of HR8799b and c}. Percentage of the orbits allowed by our analysis of the current astrometric history of the HR8799 system which preclude a ``close encounter'' with the next orbital period of HR8799b as a function of each planet's masses.} 
\label{fig:Bounds} 
\end{figure*}

Our results are illustrated on Figure \ref{fig:Bounds}, where we display the likelihood of orbits compatible with the astrometric history of HR8799 without close encounters, assuming the posterior distributions discussed in Section \ref{sec:Astro}. The top right of the two rightmost panels shows that none of the orbits estimated by our analysis of the astrometric history of HR8799 yields dynamically stable orbits for masses $>60 \; M_{Jup}$. The bottom left indicates on the contrary the region for which more than $68 \%$ of the combinations of our allowable orbits does not feature a close encounter (indicated by a a red dashed line).

Note that this boundary is arbitrary and only shown for illustration purposes. Indeed only one sigle stable orbital configuration (e.g. $10^{-4} \%$ in our case) suffices to yield an orbital architecture without a close encounter. Indeed, no rigorous dynamical mass bounds can be firmly established for HR8799 b-c using Figure \ref{fig:Bounds} since the large scatter in orbital elements resulting from our orbital motion analysis can lead to artificially low masses. However this figure provides an indication of how dynamical considerations will be able to further constrain orbital parameters and masses. Note moreover that our close encounter criterion is extremely loose when it comes to constraining dynamical masses: the distance boundary defined \citet{cfm08} is designed to rule out orbital nearby approaches which will yield to ejections in the next $10^{5}$ years. Clearly more subtle dynamical interactions can occur on scales larger than four mutual Hill radii. Moreover the duration of our calculation is conservatively limited to twice the period of HR8799b (we assume that in such short time scales the secular perturbations of orbital elements are negligible), which is clearly not sufficient to explore all the possible geometries. 

On the other hand we find that all of the orbit pair underlying the leftmost panel of Figure \ref{fig:Bounds} feature a close encounter if the masses of HR8799d-e are above $12 M_{Jup}$. While $68 \%$ boundary might be artificially low due to orbital elements scatter, the lack of stable orbits above $12 M_{Jup}$ indicates that {\bf it is very unlikely that orbits matching the astrometric history of this system will yield dynamically stable configurations unless the masses of at least HR8799de are below $13 \; M_{Jup}$} (unless there remain unidentified pathological biases in our Bayesian analysis). Note that here we adopt $m<13 M_{Jup}$ as ``planetary mass'' for the sake of the argument, in spite of the boundary in the planet/brown dwarf classification being truly tied to formation history, and thus still an active area of investigation. This is, to our knowledge, the most ``assumption free'' constraint on the dynamical masses in the HR8799 planetary system, since our analysis only relies on the very weak assumptions regarding the {\em uncorrelated} prior distributions of Keplerian elements for each planet. This approach complements the current literature which focused on identifying dynamically stable configurations while assuming coplanarity and/or circular orbits \citep{fm10,2013arXiv1308.6462G,2013A&A...549A..52E,rks09}. In a future communication we will substitute this loose close encounter criterion by full N-body simulations and thus merge both approaches in order to shed further light on the orbital architectures of this planetary system \citep{AArronDynamocal}.

\section{Conclusion}
In this paper we have presented a new astrometric perspective on the the four sub-stellar objects orbiting HR8799. We relied on the same Project 1640 observations that lead to a parent publication focused on their spectral characterization \citep{2013ApJ...768...24O}. In that paper we demonstrated how the combination of state of the art coronagraphs and adaptive optics systems with an Integral Field Spectrograph could provide tremendous insights on the atmospheric diversity of such purported planets. Herein we first focused on the intricacies associated with astrometric estimation using such a complex system. In particular we introduced two new algorithms, which respectively retrieve the stellar focal plane position (even when masked by a coronagraphic stop), and yield precise astrometry and spectro-photometry of faint point sources even when they are initially buried in the speckle noise. This latter algorithm was built upon a recent publication by \citet{2012ApJ...755L..28S} and is now becoming a standard tool in the field of high contrast imaging. The Principal Component Analysis underlying our KLIP algorithm can moreover be furthered to capture the true three dimensional stochastic nature of the speckles in IFU data, as demonstrated in a future publication reporting a novel method also developed by the P1640 team \citep{s4}. We hope that our detailed discussions regarding the intricacies of incoherent speckle suppression in high contrast IFU's, and the tenuous tasks of estimating stellar location in the presence of a coronagraphs, will facilitate upcoming large scale surveys.

The second part of our paper was devoted to the interpretation of the published astrometric history of the HR8799, augmented with our Project 1640 epoch. In order to complement the various interpretations currently in the literature we conducted a Bayesian analysis based on Markov Chain Monte Carlo, using the methods described in \citet{2006ApJ...642..505F,2012A&A...542A..41C,2013ApJ...775...56K}. Under the caveats associated with the sensitivity of such method to unaccounted astrometric biases, we were able to determine an ensemble of likely Keplerian orbits for HR8799bcde without any prior assumptions on the overall configuration of the system. We then discussed the implications of our results in terms of orbital architecture, that can be summarized below:
\begin{itemize}
\item The four planets appear to be coplanar in the broad sense of our outer solar system. However HR8799d orbits slightly outside of the plane of HR8799bce, with a misalignment in relative inclination of $\sim 15^{\circ}$. More data is necessary to unambiguously rule out coplanarity. 
\item It is particularly interesting to note that planet d both appears to have a different inclination, and eccentricity. If confirmed in the future with additional orbital data, this result would be particularly interesting to help understand the history of the system dynamic, where some event might have pumped both the eccentricity and inclination in a system otherwise mostly circularized and coplanar. 
\item The majority of recent publications discussing the orbital architecture and dynamical stability of the HR8799 system assumed strict coplanarity for the four planets. Our results raises questions about the validity of this assumption and as a consequence yields updated eccentricities hierarchy and most likely period ratios different than previously thought. 
\item Based on the set of most likely orbits established by our analysis of the astrometric history and a loose dynamical survival argument based on geometric close encounters, we have established a very high likelihood of masses below $13 \; M_{Jup}$ for HR8799de and illustrated how future dynamical analyses will further constrain dynamical masses in the entire system. 
\end{itemize}
%In an upcoming publication we will propagate our ensemble of likely orbits through N-body simulations in order to further constrain this likely subset of configurations to the few architecture that both are favored by the data. 
In an upcoming publication, we will propagate our ensemble of likely orbits through N-body simulations in order to further constrain this likely subset of configurations to the few architecture that both are favored by the data. This effort will eventually provide robust dynamical mass estimates and, when combined with our low resolution spectroscopic observations will provide a critical piece to the current puzzle associated with the formation history of this system.

\acknowledgments
A portion of this work is or was supported by the National Science Foundation under Award Nos.~AST-0215793, 0334916, 0520822, 0619922, 0804417, 0908497, 1039790 and EAGER grant 1245018. A portion of the research in this paper was carried out at the Jet Propulsion Laboratory, California Institute of Technology, under a contract with the National Aeronautics and Space Administration and was funded by internal Research and Technology Development funds. A portion of this work was supported by NASA Origins of the Solar System Grant No.~NMO7100830/102190, and NASA APRA grant No.~08-APRA08-0117. BRO acknowledges continued support from Paco. Our team is also grateful to the Plymouth Hill Foundation, and an anonymous donor, as well as the efforts of Mike Werner, Paul Goldsmith, Jacob van Zyl, and Stephanie Hunt. ELR acknowledges support from NASA through the American Astronomical Society's Small Research Grant Program. RN performed this work with funding through a grant from Helge Ax:son Johnson's foundation. LP performed this work in part under contract with the California Institute of Technology funded by NASA through the Sagan Fellowship Program. SH was supported by a National Science Foundation Astronomy and Astrophysics Postdoctoral Fellowship under Award No.~AST-1203023. Any opinions, findings, and conclusions or recommendations expressed in this material are those of the authors and do not necessarily reflect the views of the National Science Foundation. We thank the the Raymond and Beverly Sacker Foundation whose generous donation allowed the purchase of the original Project 1640 detector. We also thank Teledyne Imaging Sensors for their help and support throughout this project. We thank the dedication and assistance of Andy Boden, Shrinivas Kulkarni and Anna Marie Hetman at Caltech Optical Observatories. Finally, the entire team expresses sincere gratitude and appreciation for the hard work of the Palomar mountain crew, especially by Bruce Baker, Mike Doyle, Carolyn Heffner, John Henning, Greg van Idsinga, Steve Kunsman, Dan McKenna, Jean Mueller, Kajsa Peffer, Kevin Rykowski, and Pam Thompson. This project would be impossible without the flexibility, responsiveness and dedication of such an effective and motivated staff.

\end{document}